\pdfoutput=1
\documentclass[sigconf]{acmart}
\usepackage{balance}
\usepackage{graphicx}
\usepackage{enumitem}
\usepackage[skins]{tcolorbox}
\usepackage{dblfloatfix}  
\usepackage{colortbl}
\usepackage{arydshln}
\usepackage{times}
\usepackage{rotating}
\usepackage{makecell}
\usepackage{tabularx}
\usepackage{booktabs}
\usepackage{wrapfig}
\usepackage{tikz}
\usetikzlibrary{angles}
\usepackage{makecell}
\usepackage{tabu}
\usepackage{multirow}
\usepackage{hyperref}
\usepackage{framed} 
\usepackage{newtxmath,amsmath}
\usepackage[framemethod=tikz]{mdframed}
\usetikzlibrary{shadows}
\usepackage{graphics}
\newmdenv[
tikzsetting= {fill=gray!10},
linewidth=1pt,
roundcorner=2pt,
shadow=false
]{myshadowbox}

\usepackage{pifont}% http://ctan.org/pkg/pifont

\makeatletter
\let\th@plain\relax
\makeatother

\hypersetup{
    linkcolor=blue,
    filecolor=magenta,      
    urlcolor=cyan,
}

\setlist[itemize]{leftmargin=0.4cm}
\setlist[enumerate]{leftmargin=0.4cm}

\newcommand{\bi}{\begin{itemize}}
\newcommand{\ei}{\end{itemize}}
\newcommand{\be}{\begin{enumerate}}
\newcommand{\ee}{\end{enumerate}}
\newcommand{\fig}[1]{Figure~\ref{fig:#1}}

\newenvironment{RQ}{\vspace{1mm}\begin{tcolorbox}[enhanced,width=3.4in,size=fbox,colback=red!5!white,drop shadow southeast,sharp corners]}{\end{tcolorbox}}

\usepackage{url}
\tikzstyle{thmbox} = [rectangle, rounded corners, draw=black, fill=gray!10]

\acmConference[ESEC/FSE 2020]{The 28th ACM Joint European Software Engineering Conference and Symposium on the Foundations of Software Engineering}{8 - 13 November, 2020}{Sacramento, California, United States}

\title[Changing  Nature of CS Software]{The Changing Nature of Computational Science Software}
\author{Huy Tu, Rishabh Agrawal, Tim Menzies}
\affiliation{ Computer Science, NC State, USA} \email{hqtu@ncsu.edu,  ragrawa3@ncsu.edu, timm@ieee.org}
\date{December 2019}

\begin{abstract}
How should software engineering be
 adapted for Computational Science (CS)?
If we understood that,
then we could better support software sustainability, verifiability, reproducibility, comprehension, and usability for CS community.
For example, improving the 
maintainability of the CS code could lead to:
(a) faster adaptation of scientific project simulations to new and efficient hardware (multi-core and heterogeneous systems); (b) better support  for larger teams to co-ordinate (through integration with interdisciplinary teams); and (c) an extended capability to  model complex phenomena.

In order to better understand computational science, this paper uses quantitative evidence (from 59 CS projects in Github) 
to check
 13 published beliefs about  CS. These beliefs reflect on
  (a)  the
nature of scientific challenges; (b) the implications of limitations of computer hardware; and (c) the cultural environment of scientific software development. 
What we found was, using this new data from Github, 
only a minority of those older beliefs  
can be endorsed.
More than half of the pre-existing beliefs are dubious, which leads us to conclude that the
nature of CS software development is
changing. 

Further, going forward, this has implications for
(1) what kinds of tools
we would propose to better support computational science and (2) research directions for both communities.

\end{abstract}
 

\begin{document}

\maketitle

\section{Introduction}

Computational Science (hereafter, CS)
field studies and develops software   to explore
 astronomy, astrophysics, chemistry, economics, genomics, molecular biology, oceanography, physics, political science,  and many   engineering fields 
There is an increasing reliance of computational methods software for science. For instance, a Nobel Prize in 2013 went to chemists using computer models to explore chemical reactions during photosynthesis. In the press release of the award, the Nobel Prize committee wrote:

\begin{quote}
{\em Today the computer is just as important a tool for chemists as the test tube \cite{nobel_2013}.}
\end{quote}

Computational scientists explore software models than manually explore the physical effects they represent because it is done in real-time, more precise, faster, cheaper, and safer.  For instance, in material science, CS explores the properties
of new materials by synthesizing them, which is very expensive so standard practice is to use software to determine
those properties (e.g. via a finite element analysis). This, in turn, enables (e.g.) the faster transition of new materials to industry. Moreover, scientific software have important and widespread impacts on our society. Specifically, in weather forecasting, predictions generated from CS
software can tell the estimated path of hurricanes. This, in turn,
allows (e.g.) affected homeowners to better protect themselves from
damaging winds. 

%From these examples, there are multiple good reasons for increasing reliance on computational methods software for science such as it is real-time, more precise, faster, cheaper, and safer to explore software models than manually explore the physical effects they represent. For instance, CS software can explore the effects of several 
%hurricanes 
%scenarios and nuclear reactions without risk to human life
%or property~\cite{heaton15_lit}.

There is much demand for better software engineering (SE) methods
for CS. For example, an investigation of the 
quality of scientific software during the ``Climategate'' scandal \cite{merali10_error} found little to no reproducibility of CS results. Improving the verifiability and maintenance of CS code would hence increase the credibility of CS results and implications. Table \ref{tab:characteristics}
lists some of the prior results
where empirical software
engineering researchers have explored computational science
(this table comes from the work of
Carver, Heaton, Basili, and Johanson \cite{carver13_perception, carver07_environment, basili08_hpc, heaton15_lit, johan18_secs}, and others).
Johanson et al. \cite{johan18_secs}   argues that SE practices will only be integrated into CS when
those practices take advantage of
the   13 beliefs  of
Table~\ref{tab:characteristics}.

\definecolor{carnationpink}{rgb}{1.0, 0.65, 0.79}
\newcommand{\DOUBT}{\colorbox{carnationpink}{{\bf Doubt}}}

\definecolor{celadon}{rgb}{0.67, 0.88, 0.69}

\newcommand{\ENDORSE}{\colorbox{celadon}{{\bf Endorse}}}

\begin{table*}[t]
\centering
\caption{Thirteen beliefs from
prior studies about Computational Science. From Johanson et al. \cite{johan18_secs}. These beliefs
divide into the three categories shown in the left-hand column.
In the far right column,
anything marked as ``no evidence''
refers to beliefs we could
not check using our Github data.}
\resizebox{0.85\linewidth}{!}{
%\begingroup\setlength{\fboxsep}{2pt}
%\colorbox{lightgray}
%
\begin{tabular}{c|l|l|c}
Category  & Characteristics & Citations & Conclusion \\ \hline

\multirow{3}{*}{\begin{tabular}[l]{@{}c@{}c@{}} 1. nature of \\ scientific  \\ challenge\end{tabular}}  &  \multirow{3}{*}{\begin{tabular}[c]{@{}l@{}l@{}} a) Requirements are Not Known up Front 
 \\ b) Verification and Validation are Difficult and Strictly Scientific 
 \\ c) Overly Formal Software Processes Restrict Research 
\end{tabular}}
& \cite{segal08_ss, carver07_environment, segal05_ss, basili08_hpc, easterbrook_cs} & \ENDORSE \\
& & \cite{carver07_environment, kanewala13_testing, carver06_hpc, Prabhu11_cssurvey, basili08_hpc} & \ENDORSE  \\
& & \cite{easterbrook_cs, segal07_problem, carver07_environment, segal08_ss} & No-Evidence \\ \hline

\multirow{4}{*}{\begin{tabular}[l]{@{}c@{}c@{}c@{}} 2. limitations \\ of computer \\ hardware \end{tabular}}  &  \multirow{4}{*}{\begin{tabular}[c]{@{}l@{}l@{}l@{}} a) Development is Driven and Limited by Hardware & 
 \\  b) Use of ``Old'' Programming Languages and Technologies 
 \\ c) Intermingling of Domain Logic and Implementation Details 
 \\ d) Conflicting Software Quality Requirements 
\end{tabular}} 
 & \cite{easterbrook_cs, faulk09_secs} & No Evidence \\
 & & \cite{basili08_hpc, carver07_environment, Prabhu11_cssurvey, kendall05_C, ragan14_pythoncs} & \DOUBT \\
 & & \cite{faulk09_secs} & \ENDORSE \\
 & & \cite{carver07_environment, basili08_hpc, carver06_hpc} & No Evidence \\ \hline

\multirow{6}{*}{\begin{tabular}[l]{@{}c@{}c@{}c@{}c@{}c@{}} 3. limitations \\ of cultural \\ differences \end{tabular}}  &  \multirow{6}{*}{\begin{tabular}[c]{@{}l@{}l@{}l@{}l@{}l@{}} a) Different Terminology 
 \\ b) Creating a Shared Understanding of a ``Code'' is Difficult 
 \\ c) Little Code Reuse 
 \\ d) Scientific Software in Itself has No Value But Still It is Long-Lived 
 \\ e) Few Scientists are Trained in Software Engineering
 \\ f) Disregard of Most Modern Software Engineering Methods 
\end{tabular}} 
 &  \cite{faulk09_secs, easterbrook_cs, boyle09_lessons} & \ENDORSE \\
 & & \cite{segal07_problem, carver06_hpc, Shull05_parallel, sanders08_risk} & \DOUBT \\
 & & \cite{Prabhu11_cssurvey, segal07_problem, basili08_hpc, carver06_hpc} & \DOUBT \\
 & & \cite{faulk09_secs, segal07_enduser, easterbrook_cs, boyle09_lessons} & \DOUBT \\
 & & \cite{segal07_enduser, basili08_hpc, carver13_perception, easterbrook_cs, sanders08_risk} & \DOUBT \\ 
\end{tabular}}
\label{tab:characteristics}
\end{table*}

Just  because  prior research endorsed
 ``X'' does not mean that  ``X'' is relevant in the current context. There are numerous examples of long-held beliefs which, on re-evaluated, proved to be incomplete or outdated~\cite{menzies17,dev16}. 
Given that, and the prominence  of 
computational science, it is 
well past time for a second look at the beliefs of 
Table~\ref{tab:characteristics}. 

A recent trend is that CS researchers store their code on open source repositories (such as Github).
Our study of the 13 beliefs mines the code and comments of dozens of the repositories of  those CS projects. Three of those beliefs cannot be explored using the data available in Github. For the remaining:
\bi
\item Assuming each belief held,
we described what effect   we would expect to see in project data,
\item Then we check if that effect actually exists in the data.
If so, then  we  {\em endorse} that belief. Otherwise, we have cause to {\em doubt} it. 
\ei

Based on the analysis of 59 CS projects, our findings and contributions include: 
\be
\item Contrary to prior research, only small number of proposed beliefs in \cite{johan18_secs} are endorsed. As discussed at the end of this paper, this has implications for the research practices and what kinds of tools we would propose to better support CS. 
\pagebreak
\item The relevance of the scientific software development beliefs may change according to time.
In this regard, it is apropos to note that
  much of the prior analysis that leads to Table \ref{tab:characteristics} was qualitative in nature (i.e. impossible to reproduce, check, or refute). This work, on the other hand, is quantitative in nature. Hence, it can be be reproduced/improved or even refuted when.  To assist in that process,  we have posted all our data and scripts at
\url{https://github.com/se4cs/se4cs}. 
\ee

The rest of this paper is structured as follows.
The next section offers some preliminary notes on the data
we collected and our methods for labelling, then analyzing,
that data. Then \S3 discusses the general threats to validity of our work. \S4-6 provide background, analysis results for much evidence
of the changing nature of computational science software. \S7-8 summarize, conclude, and offer future directions of SE for CS research.

\section{Preliminaries} 

%The growing dependency of science on computational methods software to make decisions for scientists is inevitable. By enhancing the software's quality, scientists can guarantee the CS work more credible and more productive. Therefore, 
%better SE improves computational
%science software, which would lead to better (e.g.) weather prediction and the faster creation of new industries based on new materials.

\subsection{Modelling Assumptions (and ``Indicators'')}\label{model}

The reasoning of this 
paper makes
modeling assumptions in order to bridge between the higher-level concept of the belief to what are measurable through the Github data. For example, consider the belief 1.b ``Verification and validation in software development for CS are difficult''.  
Having read 10,000s of comments, we can assert that 
very few commits are labelled  ``verification and validation'' (V\&V) and, of those that are,
even less use these terms in a consistent manner. Instead, based on our reading of the commits, we could assign labels showing whether or not developers were reporting the results of creating/running tests. Hence, to explore that belief we had to make the following modeling assumptions to bridge between the terminology of the belief and the terms in the Github data: (1) V\&V is associated with testing; and (2) the amount of testing is an indicator for V\&V activity. This modelling assumption that relies on commits to indicate the amount of developers effort in a specific  task is also done by other SE researchers \cite{vasilescu16_limit, xia2019sequential}.

Formally, this means that our conclusions are based on what
Schouten et al.describe as
 {\em indicators}~\cite{schouten2010indicators} rather than direct measures. Indicator-based reasoning is often used as a method to take steps closer to intangible/ abstract/ expensive vision. For example,
 in statistics, Schouten relied on indicators to support large survey data collection monitoring \cite{schouten2010indicators}. Also, in SE, Lamsweerde used indicators to evaluate the degree of fulfillment of goals \cite{vanLamsweerde2009_requirement}.
 Further, in business 
 management, Kaplan and Norton \cite{kaplan1996using} offered a four-layer ``perspectives diagram'' that implements the bridge from high-level and intangible business goals down
 to observable entities, i.e. indicators (at the time of this
 writing, that paper has  9800+ citations in Google Scholar).

% \subsection{Research Questions}

% There have been many claims and many studies made about how scientists develop software along with systematic reviews of the literature from both scientists and SE community about developing software. It is essential to not only survey literatures of both community but also conducting quantitative study and interviews to validate the 13 characteristics:

% \begin{enumerate}
%     \item What claims have scientists indicated within the characteristics/constraints about why state of the art SE techniques are poorly adopted by scientists?
%     \item What empirical evidence (e.g. quantitative study of the project development or interviews) to validate or reject these claims? 
% \end{enumerate}

\subsection{Data Collection}\label{tion:data}

To check our beliefs on CS projects, we proceeded as follows. 
Using our contacts in the CS community
(from the Molecular Sciences Software Institute (MOLSSI), and the Science Gateways Community Institute (SGCI)) we found
678 CS  projects.
Researchers
warn against using all the Github data~\cite{bird09promise,agrawal2018we, eirini15promise, munaiah17curating} since
many of these projects are simple one-person prototypes.
Following  their advice, we applied the sanity checks of Table \ref{tbl:sanity}
to select 59  projects with sufficient software development information
(for space reasons, we list those projects outside of this paper in our on-line materials; see \url{https://github.com/se4cs/se4cs}).
 
\definecolor{amethyst}{rgb}{0.6, 0.4, 0.8}\definecolor{cadetblue}{rgb}{0.37, 0.62, 0.63}
\begin{figure*}[!t]
\vspace{5pt}
\centering \includegraphics[width=.975\linewidth]{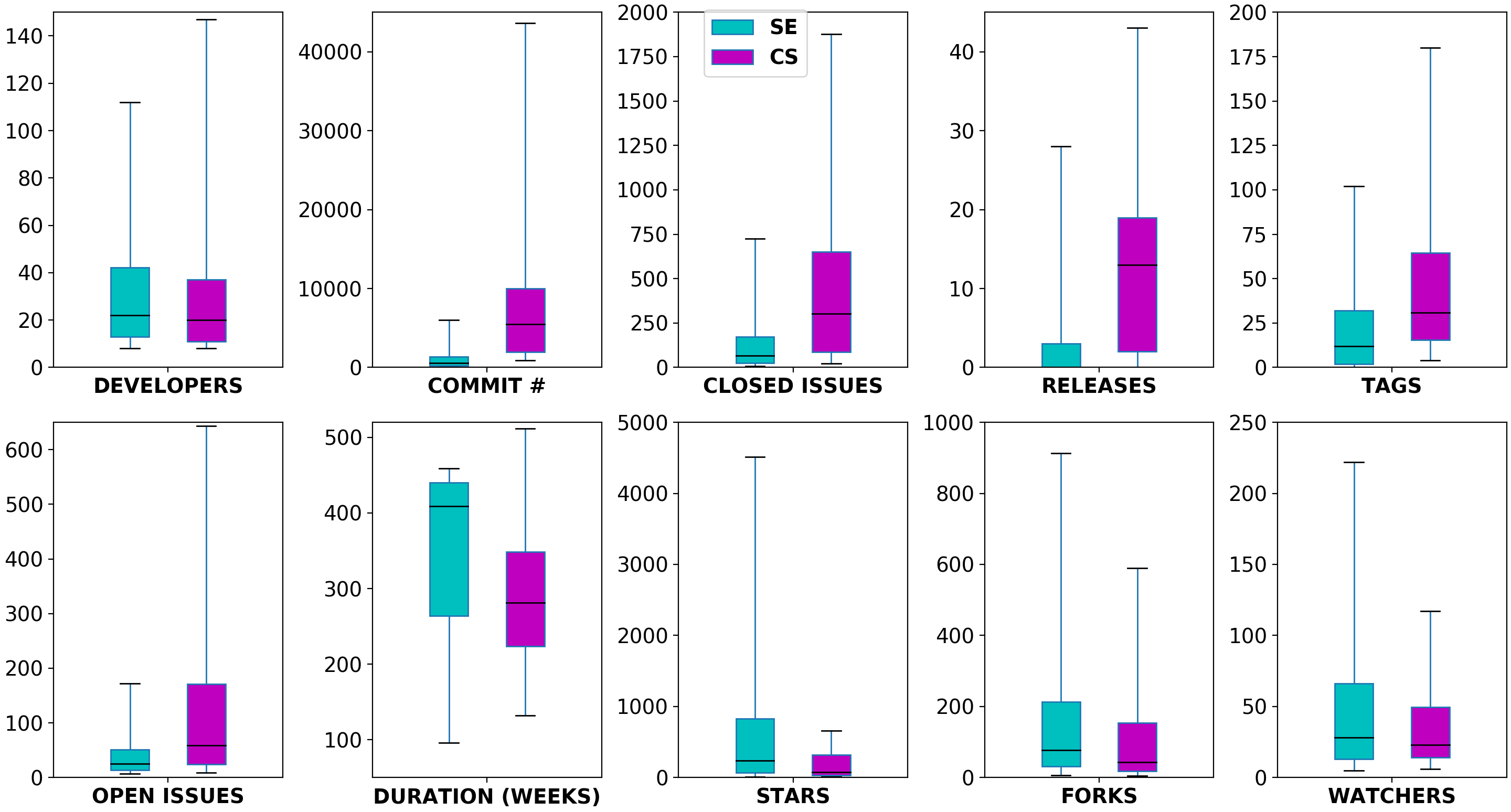}
\caption{Data distributions from 1300 SE projects (shown in \colorbox{cadetblue}{ \textcolor{white}{teal}}) \& 59 CS projects (shown in \colorbox{amethyst}{ \textcolor{white}{purple}}).}\label{fig:comparison}
\end{figure*}

\begin{table}[!t]
\caption{Data sanity checks. From \cite{Kalliamvakou:2014}.}\label{tbl:sanity}
\small
 
%{\small
 \begin{tabular}{r|l}
 Check   & Condition    \\\hline
 \# Developers & $\geq$ 7 \\
 Pull requests  & $>$ 0 \\
Issues & $>$ 10 \\
Releases &  $>$ 1 \\
Commits & $>$ 20 \\
Duration  & $>$ 1 year 
\end{tabular}%}
 
\end{table}
Figure~\ref{fig:comparison} shows some statistics on the data we collected from our 59 CS projects. For comparison purposes, we compare that sample to 
a sample of 1037 Github projects from~\cite{Majumder19}.
There is no overlap between the CS projects and the Github sample. Also, all the Github
sample passes the sanity checks of Table \ref{tbl:sanity}. Figure~\ref{fig:comparison} uses the following terminology.

\textit{Developers}: Developers are the contributors to a project, who code and submit their code using commit to the code base. The number of developers signifies the interest of developers in actively participating in the project and volume of the work.

\textit{Commits:} in version control systems, a commit adds the latest changes to [part of] the source code to the repository, making these changes part of the head revision of the repository. 

\textit{Open \& Closed Issues:} Users and developers of a repository on Github use issues as a place to track ideas, enhancements, tasks, or bugs for work. As they work, they open issues with Github. When developers address those matters, they close the issues.

\textit{Tags}: Tags are references that point to a specific time in the Git version control history. Tagging is generally used for marking version release (i.e. v1.0.1).

\textit{Releases:} Releases mark a specific point in the repository’s history. The number of releases defines different versions published (and  signifies a considerable  changes  between each version).

\textit{Duration:} The duration of a project marks the length of the project from its inception to the current date or project archive date (in week as a unit of time). It signifies how long a project has been running and in the active development phase.

\textit{Stars:} A repository's stars signify how many people
  ``liked'' a project enough to create a bookmark to follow its future progress.
  
   \textit{Forks}: A fork is a copy of a repository. Forking a repository allows users to freely experiment with changes without affecting the original project. This number
  is an indicator of how many people are interested in the repository and actively thinking
  of modification of the original version.
  
   \textit{Watchers}: Watchers are GitHub users who have asked to be notified of activity in a repository, but have not become collaborators. This is a representative of people actively monitoring projects, because of possible interest or dependency.

  The following observation  
  will become important, later in the paper (i.e. issues of size conflation).
Assuming that ``standard'' SE projects are those we see in Github which pass the sanity checks of Table \ref{tbl:sanity}, then
 Figure~\ref{fig:comparison} shows that
  \begin{quote}
  {\em It is not true that CS projects are usually smaller than standard SE projects.}
  \end{quote}
 To justify this statement, we applied 
  a 95\% confidence bootstrap statistical test~\cite{efron94} and an A12 effect size test~\cite{arcuri2011practical}, to all the  Figure~\ref{fig:comparison} distributions where the median CS values were lower than the median SE values.
Only in the case of {\em duration} were the median CS values statistically different and less than the SE medians. All the other indicators show that CS projects are just as active (or even more active) that SE projects 

The one clear ``less than'' result of Figure~\ref{fig:comparison} is that the {\em duration} of the CS projects is less than that of the SE projects (281 weeks versus 409 weeks). 
This is interesting since it suggests
that the CS developer community is working
just as hard (or even header)
as the SE communuty, {\em and does so in less time}.
This suggests that SE has more to learn from CS than the other way around.
If we say that   an {\em efficient} software process is one that allows  people to work together, faster,  then 
Figure~\ref{fig:comparison} is saying:
\begin{quote}
{\em CS software development is  more efficient that SE.}
\end{quote}

 \subsection{Labelling}
 When code is shared
within a software repository, an important event is the {\em commit comments}. These comments are the remarks developers make to document and justify some updates to the code, i.e. a rich source of information about a project. Code repository systems such as Github store tens of millions of these comments that are utilized as a rich source of information about a project within SE literature. For instance, within SE literature,  Vasilescu et. al \cite{vasilescu16_limit} and Menzies et. al \cite{xia2019sequential} studied commits as an indicator for development effort of projects.

To understand the scientific development process, we manually categorized the commit comments seen within CS
software. 
Using the power of free pizza, we assembled a team of 10 computer science 
graduate students.
To allow other researchers to reproduce this work, we set
the following
resource limit on our analysis.
According to Tu et al.~\cite{tu2019better}, two humans can manually read and categorize
and cross-check 400 commit comments per day (on average).
Hence, for this study, for each project, we categorized 400 commits
(selected at random). 
All in all, our
reviewer team spent 320 hours (in total) categorizing comments.

 Our  reviewers
labeled commits using the following
guidelines:
\bi
\item {\em Science enhancement}: any core science (e.g. an equation of Pascal triangle) that is being implemented or modified.
\item {\em Engineering enhancement:} any other enhancements that related to code complexity (e.g. data structures \& types, I/O formats, etc) 
\item {\em Bug fixes:} Fixing software faults reported or found within the development. 
\item {\em Testing: } evaluate the functionality of a software application (e.g. scientific calculations to output/input formats).
\item
{\em Other:} not core changes, e.g. renaming or formatting changes
\ei

Each commit was labeled by two reviewers,
neither of which had access to the other's labels. Moreover, the reviewers did not only look at the commit message but also the code contribution associated with the commit (e.g. to determine if the nature of some enhancement was 
``scientific'' or ``engineering'' in nature). The level of
labeling disagreement was low (just 19\%). When labels disagreed, the commit was given to our most experienced reviewer who made an executive decision about what
was the correct label.

\subsection{Beliefs We Cannot Explore (Using Github)}

Github stores data about code and the comments seen during code reviews and pull requests. While this is useful for assessing most of the beliefs of Table~\ref{tab:characteristics}, it does mean that at least three of the thirteen beliefs, summarized by Johanson et al. \cite{johan18_secs}, cannot be explored by this paper:

\be
\item {\em Overly Formal Software Processes Restrict Research}: Computational scientists perform many tasks,
only one of which is developing software. For example,
they must write grants, do presentations, traveling, keeping up with the fast-developing fields, etc. Hence, measuring the formality of software processes and research efforts would be outside of the scope for Github.
\item {\em Development is Driven and Limited by Hardware:}
We found it difficult to access information about hardware platforms from our Github data. Hence, we cannot reason about this belief.
\item {\em Conflicting Software Quality Requirements:} These requirements include functional correctness vs
performance or portability or maintainability. Specifically, performance issues conflict with portability and maintainability since these are often achieved via hardware-specific optimizations. As with issues
relating to hardware,
the information rarely exists on Github. 
\ee

\section{Threats to Validity}

\subsection{External Validity}
Like any data mining paper,
the results of the following analysis are skewed by sampling bias.
To combat that effect, when we analyze Github data, we took care to analyze as much as possible.
Hence,
as a starting point of this work, we looked at 687 CS projects. 
Using the advice from Kalliamvakou et al.~\cite{Kalliamvakou:2014}, we applied certain sanity checks of Table~\ref{tbl:sanity}  to focus on 59 of those 687 projects.

At 59 projects, this sample is much larger than seen in
most prior studies on computational
science. That said, it is certainly true that another sample of different projects would make different conclusions.
Accordingly, we make all our scripts and data publicly so that
(a)~our current conclusions
are repeatable/ refutable/ improvable can be quickly repeated across multiple projects by anyone with access to Github; and (b)~our current conclusions can be checked against other data, whenever that becomes available.

\subsection{Construct Validity}
As mentioned above, the following analysis depends on numerous {\em indicators} to bridge between the belief being explored and the available data. 
In the following,  we will take care to carefully document
the {\em modeling assumptions} used to design those indicators.

\section{Beliefs about the Nature of the Scientific Challenge}

\subsection{Requirements}\label{rments}

Our analysis of this first belief will conclude that
CS code is built in an exploratory manner,
rather than in response to some pre-defined
requirements. While this first conclusion is hardly
surprising, it does offer a simple example
of how this paper uses Github
data to reason about CS projects.

\noindent \textit{\underline{Belief:}} According to Basili (and others), in computational science,
project requirements are not known up front \cite{segal08_ss, carver07_environment, segal05_ss, basili08_hpc, easterbrook_cs}.
If true, then this belief means that SE methods based on 
static requirements (e.g. model checking) are not so valuable
for CS software.

\noindent \textit{\underline{Notes:}} 
Many authors, including Carver~\cite{carver07_environment}
and Easterbrook~\cite{easterbrook_cs}
comment that CS code is not written in order
to satisfy some pre-existing set of requirements.
Rather, it is written an exploratory fashion in order
to better understand some effects. 
This would make CS software very different to code developed using (e.g.) a waterfall model where the requirements
are all known at the start of the development.

\begin{figure*}[!b]
\begin{center}
\resizebox{1\textwidth}{!}{
 \hspace{-2mm}\includegraphics[width=.42\textwidth]{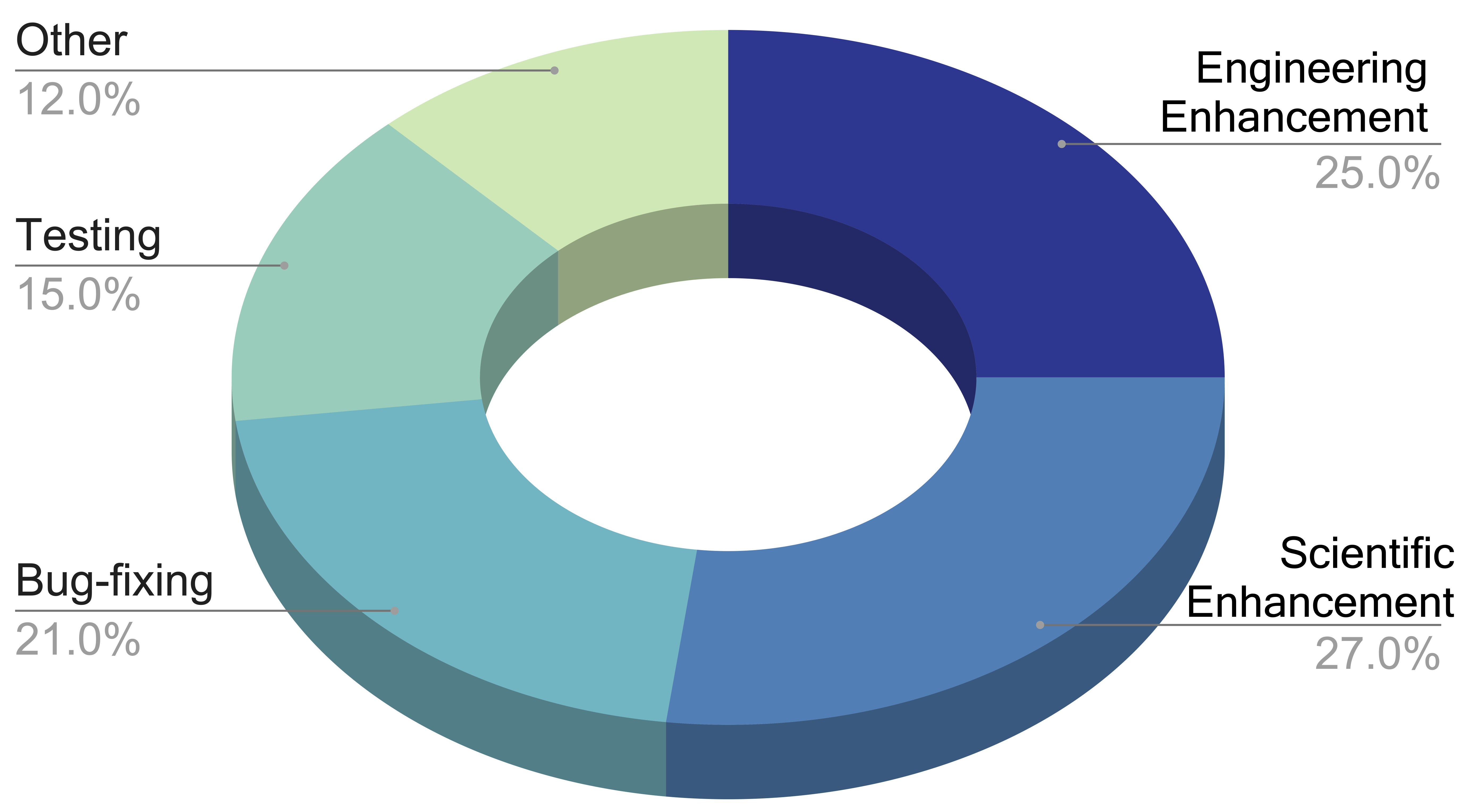}
\includegraphics[width=.47\textwidth]{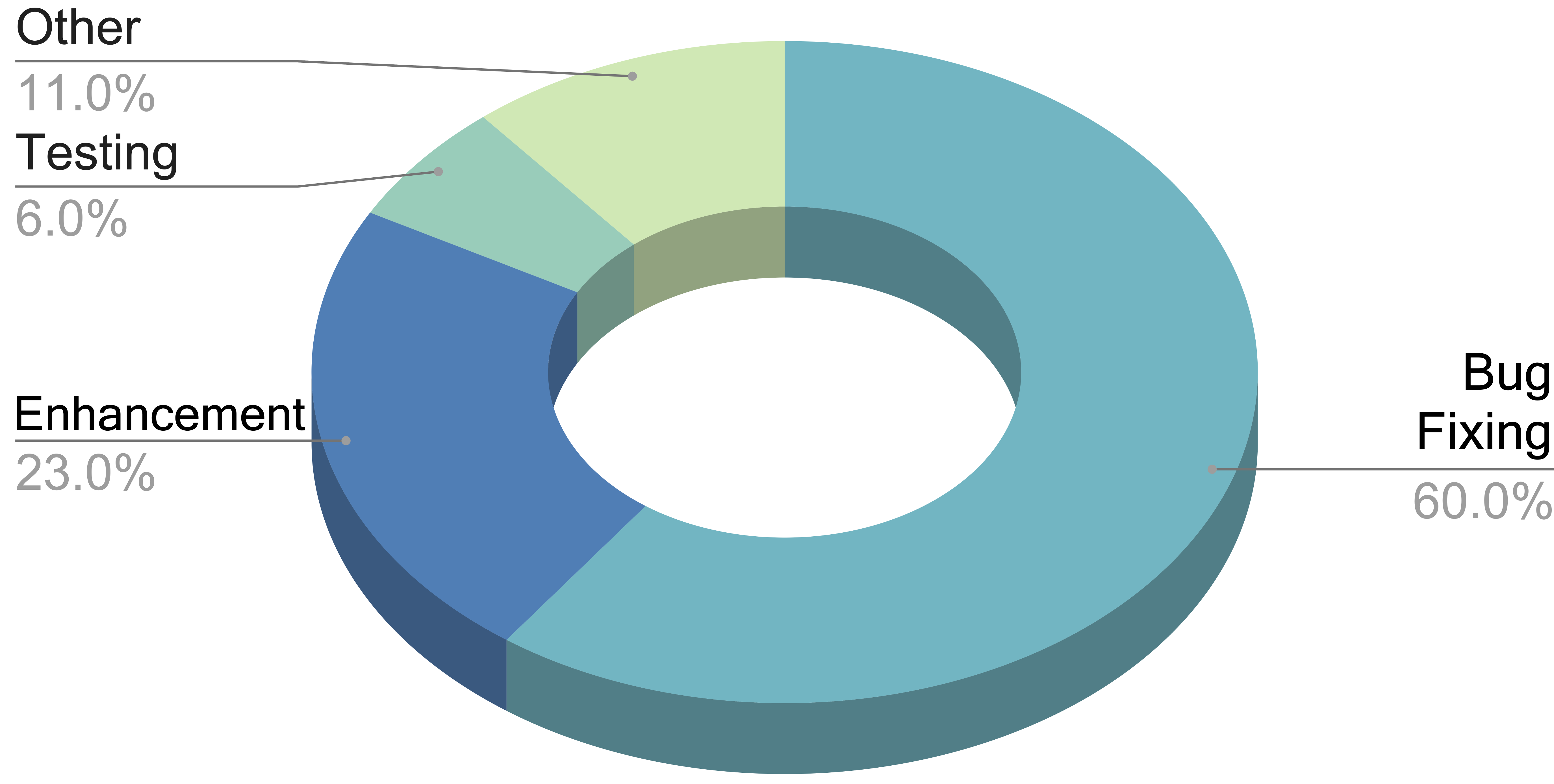}}
 
\end{center}
\caption{Distribution of development within our sample of 59 CS projects (top) and 20 top sampled SE projects (bottom).}
\label{fig:SE_activities}
\end{figure*}

\begin{figure}[!t]
\begin{center}\includegraphics[width=0.9\linewidth]{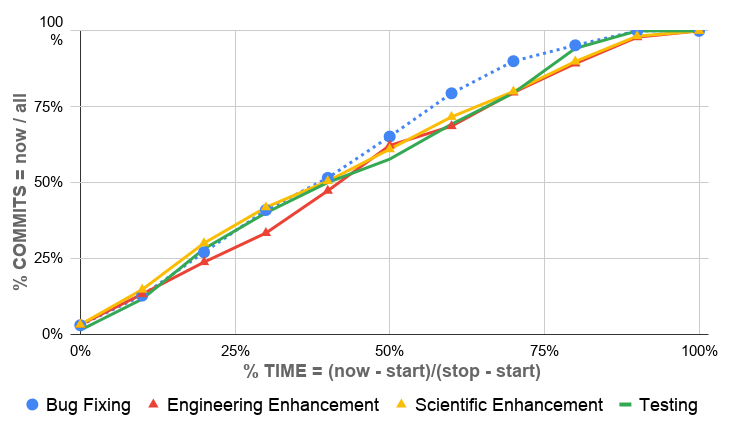}\end{center} 

\caption{Median percent  of total commits seen
at 10, 20, 30, ... 100\% of 
the time these projects were
documented in Github.
X-axis measures time as percent of days seen in Github.
Note
that all commit types occur
at a similar, and near constant,
 rate, across the lifetime of a project.}\label{fig:belief1} 
\vspace{-5mm} 
\end{figure}

\noindent \textit{\underline{Modeling Assumptions:}} 
Projects with pre-existing list of fixed requirements 
can be developed in a ``waterfall'' style.
When that style is applied,
requirements is followed by analysis,
design, code, implementation and test.
The observable feature of such projects
is that most of the testing and bug fixing
activity occurs {\em after} a code base
has been enhanced with the required
scientific or engineering functionality.

\noindent \textit{\underline{Prediction: }} If CS software was written in response to some pre-existing set of requirements, then
we would expect to see bug-fixing and testing to be a predominately end-stage activity.

\noindent \textit{\underline{Observed:}} 
As shown in \fig{belief1}, the rate
of commits of different types
is nearly constant across the project
lifetime. This observation is {\em not} consistent with 
waterfall-style projects where most of the enhancement work happens early in the lifecycle and most of the test work happens later on.
\vspace{-2mm}
\begin{RQ}
\textit{\underline{Conclusion:}}
We \textbf{endorse} the belief that, in CS, project requirements are usually not pre-defined
at the start of a project.
\end{RQ}

\subsection{Verification and Validation is Different}\label{vv}
\textit{\underline{Belief:}} 
According to Carver et al., and others, 
verification and validation in software development for CS is difficult and strictly scientific \cite{carver07_environment, kanewala13_testing, carver06_hpc, Prabhu11_cssurvey, basili08_hpc}.
That is, CS developers spend more time debugging their theories
of physical phenomena than debugging systems issues within their code.
If this belief were true then much of the standard SE testing
infrastructure would need extending before it can be applied to CS. For example,
while unit tests and system tests are certainly useful, CS projects would need a separate level of tests for ``physical concept testing''. 

\noindent \textit{\underline{Notes:}} 
According to Carver et al.~\cite{carver07_environment},
verification and validation of scientific software should be difficult for several reasons:
\bi
  \item Lack of suitable test oracles \cite{kanewala13_testing},
  \item Complex distributed hardware environments with no comparable software \cite{basili08_hpc},
  \item Scientists often suspect that the problems of the software is the results of their scientific theory~\cite{faulk09_secs},
  \item Lack of physical experimentation and experimental validation is impractical \cite{carver07_environment}. 
\ei

\begin{wraptable}{r}{1.7in}
\caption{Labels of testing type commits from the labeled Testing commits.}\label{tbl:testing}
\footnotesize \begin{tabular}{l|c|c}
\multicolumn{1}{c|}{} & \multicolumn{1}{c|}{Absolute} & \multicolumn{1}{c}{Percent}\\
\hline
Science & 289 & 47\% \\
Engineering & 146 & 24\% \\
Other & 173 & 29\% 
\end{tabular}
%}
%\end{threeparttable} 
\end{wraptable}\noindent~\textit{\underline{Modeling Assumptions:}} 
As stated above in \S\ref{model},
in order to bridge between the terminology of the belief and the Github data, we assume that (1) V\&V is associated with testing; and (2) the amount of testing is an indicator for V\&V activity. This approach is similar to studies done by Vasilescu et al. \cite{vasilescu16_limit} or Menzies et al. \cite{xia2019sequential}, where the number and the proportion of commits are treated as an indicator for developing efforts of the repositories. 

\noindent ~\textit{\underline{Prediction: }}
Verification and validation in CS is more
``difficult'' than in SE if the observed CS effort in this area
is much larger than in SE. As to ``strictly scientific'', we should see far more ``scientific
testing'' that otherwise (e.g. ``engineering testing'').

\noindent ~\textit{\underline{Observed:}}
It is easy to show that CS software verification and validation are heavily focused on scientific issues.
Table \ref{tbl:testing} shows that ``scientific testing'' is the largest type of commit in our labeled Testing commits sample (at 45\%). 
Far less effort is spent on ``engineering testing'' (only 24\%). 
As to showing the CS verification and validation is ``more difficult'' than in SE,
  Figure \ref{fig:SE_activities} shows that
  15\%, 6\% percent of the commits
  are associated with CS, SE testing (repsetively).   This SE data comes from a recent study \cite{tu2019better} of the top-20 highly starred from Github that satisfies our sanity checks of Table~\ref{tbl:sanity}.
Note that  15\% is 2.5 times larger than 6\%. That is
to say, for  verification and validation, 
much more effort is being spent in CS projects than SE.

% \bi
% \item CS software is written to correspond to physical phenomena, the nature of which may never change (e.g. the atomic weight of iron).
% \item
% the highly starred projects in Github) is written to correspond to an ever-changing ecology of platforms, tools, user expectations, and newly-arrive AI algorithms, etc.   
% Hence, it is not surprising  SE software requires more verification and validation effort than CS software since the problem it addresses are more dynamic.

This result is somewhat strange since it runs counter to standard beliefs in the SE literature (e.g. Brookes argues that unit tests and systems tests will consume half the time of any project~\cite{brooks1995mythical}). One of our conjectures include the larger V\&V effort in SE  is due to the nature of CS problems. CS software is written to correspond to solve endless nature's problems (most are beyond human's understanding) with the requirements are not known up front and software's state are incrementally improved. CS V\&V have to cover both scientific and engineering concerns while SE V\&V at some points would mature to only focusing on verification (especially when SE software is based on production focus). 

More intuitively, by looking at the \textit{Testing} and \textit{Bug-fixing} attributes from Figure \ref{fig:SE_activities}, the bug-fixing activities from SE software development are almost three times as in CS which is the direct result from testing 2.5 times less than CS. Essentially, the \textit{less} developers test, the \textit{more} bugs developers have to fix. After shipping the software, SE developers are more reluctant to test the software while for CS developers, scientific software research and development might be a continuous journey.

Moreover, a conflating factor that might make us doubt this observation would be if the CS codes were always much smaller than the SE codes. If that were true then even if some tasks had a larger percentage effort 
(e.g. Table \ref{tbl:testing}'s ``scientific testing'') then  ``relatively more'' might actually
mean ``less'' (in absolute terms). 
As discussed in \S\ref{tion:data} our data does not show that  SE projects are larger and more active
than in CS projects.

Hence, 

% Among all the defects fixing, the scientific and engineering defects are at the same rate. Yet, the testing focuses solely on the scientific aspect, almost three times (45\%/17\%), more than engineering testing. In a sense, scientists solely believe that the software is defected due to their science understanding when transferring that to source code while overlooking the engineering aspect. Yet, it is understandable because scientists have a lot of responsibilities (read and write papers, grants, give presentations, develop scientific models, etc) so they can only focus on testing on what they good at, i.e. scientific models. It is possibly useful for the community to incorporate automated SE testing tools for CS projects. 

\begin{RQ}
\textit{\underline{Conclusion:}}
We \textbf{endorse} the belief that within CS, software development's verification and validation, are difficult and mostly concerned with scientific issues. 
\end{RQ}

\section{Beliefs about Limitations of Computer Hardware}

% In this section, we discuss characteristics of software development in
% computational science that are due to limitations regarding available computing
% resources and their efficient programming. 

\subsection{Use of ``Old'' Techniques (and a Disregard for  Recent SE Methods)}\label{lang}
This section explores belief 2b (CS teams use ``old'' SE techniques)
and, as a side effect, belief 3f (CS disregards most modern SE methods).

\noindent \textit{\underline{Belief:}} According to Basili et al., and others~\cite{basili08_hpc, carver07_environment, Prabhu11_cssurvey, kendall05_C, ragan14_pythoncs},
computational scientists prefer
``older''-style programming languages and technologies while disregarding most of the newer SE methods

\noindent \textit{\underline{Notes:}} The usual argument here is that CS Scientists are skeptical of modern SE methods and new technologies/languages.
This is based on several factors: 
\begin{itemize}
  \item A decades-long commitment with these older-style languages (Fortran and C) on high-performance computing platforms \cite{faulk09_secs}.
  \item A belief that the extra features of the newer languages needlessly conflate functionality that can be more easily implemented in (e.g.) one line of ``C'' macros \cite{sanders08_risk}. 
  \item A prejudice against the never languages or a perception that the scientists would not find then useful \cite{Prabhu11_cssurvey}. 
\end{itemize}

\noindent \textit{\underline{Modeling Assumptions:}} 
One indicator of using ``new'' techniques is the presence of automatic testing and deployment tools; e.g. use of the Travis CI tool that re-runs test suites whenever new code is committed to a repository. 

Another indicator is the development language for the project. 
Johanson et al.~\cite{johan18_secs} say that, in CS, Fortran and C are examples of this ``old'' technology. The use of C++ is an interesting borderline case- Johanson et al. regard that as ``new technology'' even though it is now decades old. In the following, we will take care to examine the C++ data as a special case.

\begin{wraptable}{r}{1.5in}

\caption{Languages used in our 59 CS projects.  
}\label{tbl:language}
 \footnotesize
%\begin{threeparttable}
%\vspace{-10pt}
%\resizebox{!}{0.2\linewidth}{
%\setlength        abcolsep{10pt}
 \hspace{-3pt}\begin{tabular}{l|c|c}
 \multicolumn{1}{c|}{} & \multicolumn{1}{c|}{Count} & \multicolumn{1}{c}{Percent}\\
\hline
Other & 3 &  5\%  \\ 
Javascript	& 2 & 3\% \\ 
C &	3 & 5\% \\ 
Java	& 5 & 9\% \\ 
Fortran	& 6 & 10\% \\
C++	& 17 & 29\% \\
Python & 23 & 39\% 
\end{tabular}
%}
%\end{threeparttable}
\end{wraptable} \noindent ~\textit{\underline{Prediction: }} If CS teams are mostly focused on ``old'' technology then most of those projects would use ``old'' languages and would not use automated testing
tools like  Travis CI.

\noindent\textit{\underline{Observed:}} 
As seen in Table~\ref{tbl:language}, C and Fortran are just 15\% of our sample.
Even if we call C++ ``old'', then the ``older'' technologies of Table~\ref{tbl:language}
cover less than half the sample (44\%).
 As to other measures of ``new'', we found that  $43/59=73\%$
have active
Travis CI connections.

Hence, we say:

\begin{RQ} 
\textit{\underline{Conclusion:}} We \textbf{doubt} the belief that 
CS developers are skeptical of modern SE methods and new technologies/languages.
\end{RQ}

\noindent \textit{\underline{Discussion:}} This result is at odds
with numerous papers~\cite{basili08_hpc, carver07_environment, Prabhu11_cssurvey, kendall05_C, ragan14_pythoncs}. We explain our novel findings as follows. 
Most of the papers that endorse this view come from before the recent Silicon Valley boom. In our discussions with postdocs and Ph.D. students working on CS projects,
we found they were well aware of the salaries they might earn if they understood the popular tools used by contemporary agile software companies. 
Hence, it is perhaps not so surprising that we report here a widespread use of modern software techniques in CS.

\subsection{Cannot Separate Domain Logic and Implementation Details} 

\noindent \textit{\underline{Belief:}} 
According to Vanter et al.~\cite{faulk09_secs},
CS developers working on scientific software development do not separate high-level domain logic with
lower-level implementation details. If true, this would somewhat restrict the ability
of this community to develop general-purpose abstractions. This, in turn, would lead
to productivity issues since new applications would have to rework much of the previous
work.

\noindent \textit{\underline{Notes:}} One measure of the mature software engineering is the
use of abstraction; i.e. the ability to step back from application-specific
details to generate domain-general abstractions. Such abstract thinking is harder to do for developers that spend more time studying physical phenomena than the code used to model that phenomena

\noindent \textit{\underline{Modeling Assumptions:}} During scientific software development,
\bi
\item Domain logic addresses computational models, i.e. core science, understanding (as manifested by scientific enhancement activities). 
\item Implementation details address coding/building   tools   to solve scientific problems (as manifested by engineering enhancement). 
\ei

\noindent \textit{\underline{Prediction: }} If domain logic and implementation details are intermingled/inseparable during the development of scientific software, then both scientific and engineering enhancement contribution distribution should occur at similar frequencies.

\noindent \textit{\underline{Observed:}} Across the enhancement type commits from the sample (from Figure \ref{fig:SE_activities}), 26\% enhancement commits focusing on the core science while the rest 29\% enhancement commits focusing on the quality of the code. The absolute difference between the two types of enhancement activities is small (3\%). 

Moreover, from Figure \ref{fig:belief1}, the rate of commits of engineering and scientific enhancement activities are observed to grow synchronously across the project lifetime. Hence:

\begin{RQ}
\textit{\underline{Conclusion:}} Lacking evidence to the contrary and with some supportive indicators, 
we \textbf{endorse}  that CS developers intertwine their work
on domain logic and implementation details.
\end{RQ}

\begin{figure*}[!t]
\resizebox{1\linewidth}{!}{
 \hspace{-2mm}
 \includegraphics[width=0.5\linewidth]{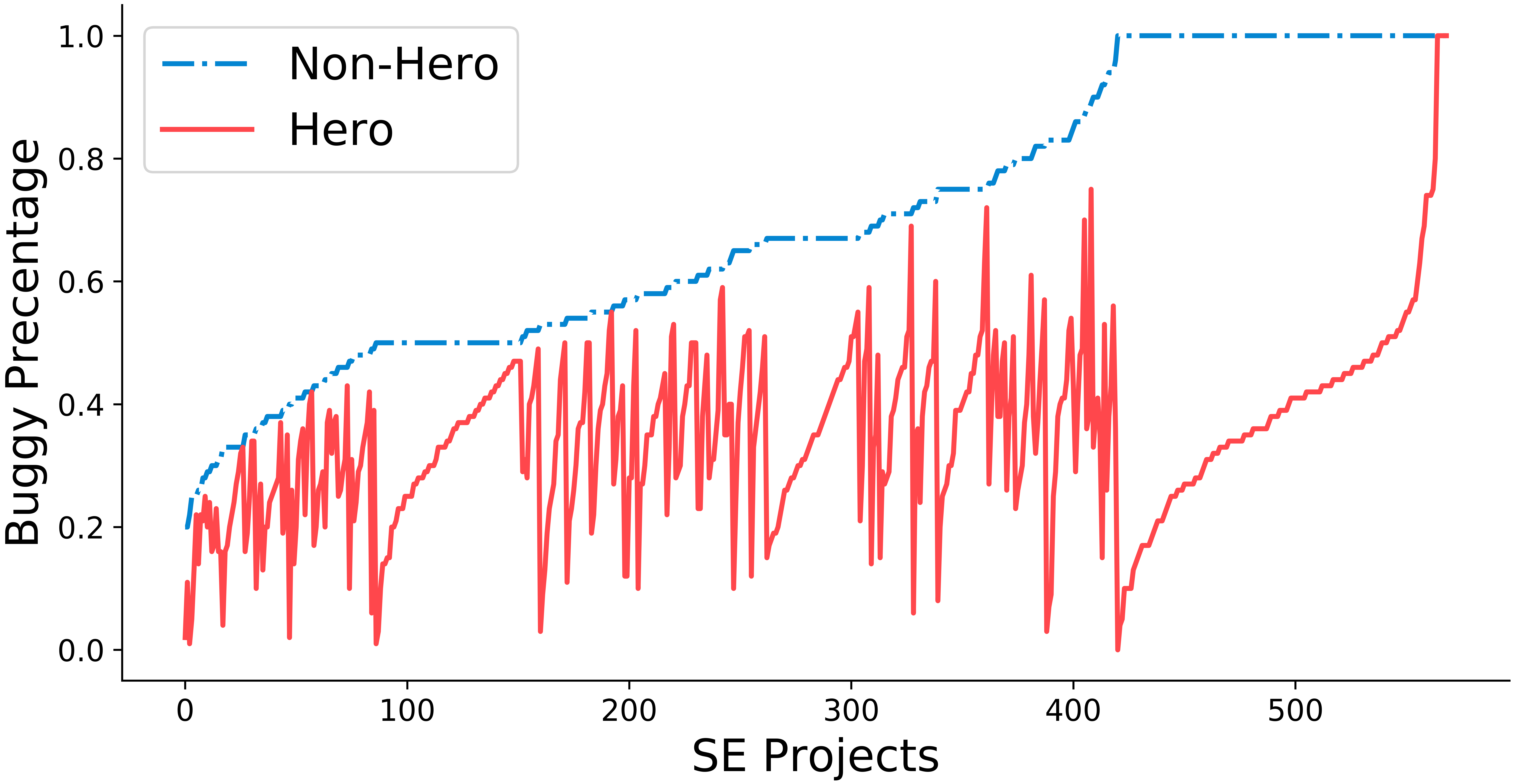}
 \hspace{2mm}
 \includegraphics[width=0.5\linewidth]{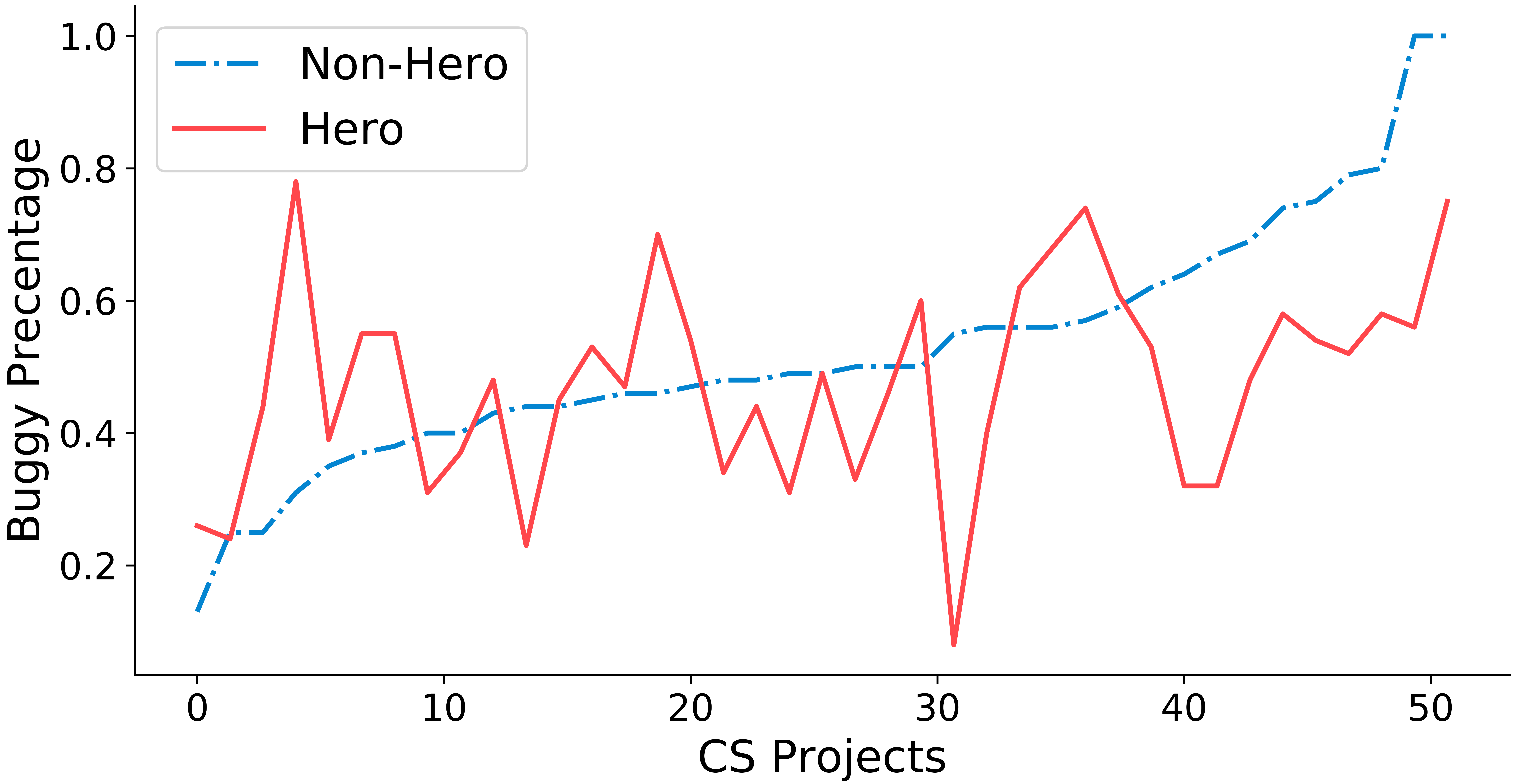}}

\caption{Percent of code commits that introduce new bugs, made by hero and non-hero developers from  SE projects (left) and CS projects (right).
Each x-axis of that figure is one project so the hero and non-hero defect introduction rate (defects per commit) is the ratio of the blue to red numbers
at any specific x-value.
}\label{fig:heroes}
\end{figure*}

\section{Beliefs about Limitations Due to Cultural Differences}

\subsection{Terminology}\label{terms}
\noindent \textit{\underline{Belief:}} 
Vanter et al.~\cite{faulk09_secs, easterbrook_cs, boyle09_lessons} express concerns
that it is hard to translate concepts between  CS and SE (since the fields are so different). 

\noindent \textit{\underline{Notes:}} When two fields evolve along different
lines (like SE and CS) it is possible that the terminology of one field has important
differences in the other field. This is worrying if those terminology differences mean that methods from one field perform poorly in the other.

\noindent \textit{\underline{Prediction: }} If the belief is held, then the off-the-shelf SE methods may perform badly of CS projects {\em unless} they first adjust the meaning of their SE terminology.

\noindent \textit{\underline{Observed:}} Tu et al. \cite{tu2019better} found that the concept of ``defective'' was  different in CS and SE.
Specifically, off-the-shelf defect labeling technologies (that are widely used  in SE \cite{tu2019better,mockus00changeskeys,kamei12_jit, hindle08_largecommits, Kim08changes}),
performed poorly when applied to CS projects.
To fix that, they built an automatic assistant called EMBLEM that showed a subject matter expert 
examples of supposedly defective CS code (as identified by the off-the-shelf SE tool).
Using feedback from the subject matter expert, the automatic assistant adjusted the support vectors of an SVM. In this way, the assistant could learner what ``defective'' means in CS.

Table~\ref{tbl:rq2aaa} 
compares defect predictions generated by a data
miner using defect labels from (a)~an off-the-shelf SE defect method;
and (b)~those generated via EMBLEM. The gray high-lighted ones are additionally added as a precaution step to check the validity of their work. 
Given $N$ releases per project and using data mining, two predictors were learned from release $i$ then tested on release $i+1$:
\bi
\item
One predictor was built using defects identified  via EMBLEM;
\item
The other predictor was built using defects identified by the standard SE defect labeler.
\ei

Note that, in most cases in  Table~\ref{tbl:rq2aaa},  EMBLEM's predictors usually out-performed the off-the-shelf SE method. 
That is, for CS projects,
better results were obtained after adjusting the meaning of a standard term (``defect'')
taken from SE.

Hence we say:

\begin{RQ}
\textit{\underline{Conclusion:}} We \textbf{endorse} that the CS community utilizes a different terminology when describing their work. SE tools
may need to be adjusted before being applied to CS projects.
\end{RQ}

\subsection{Code Understanding} ~\\
\noindent \textit{\underline{Belief:}} According to
Segal et al., and others~\cite{segal07_problem, carver06_hpc, Shull05_parallel, sanders08_risk},
CS projects are so complex that creating
a shared understanding of that code is difficult. 

\newcommand{\varendash}[1][5pt]{%
  \makebox[#1]{\leaders\hbox{--}\hfill\kern0pt}%
}

\newcommand{\RULEE}[1]{\textcolor{black!20}{\rule{#1}{6pt}}}
\begin{table}[!t]
\caption{Given $N$ releases of software, this chart shows the percent of releases
where off-the-shelf SE defect predictor is defeated by the
EMBLEM defect predictor (that learns what ``defect'' means for CS). }
\label{tbl:rq2aaa}
\footnotesize
\begin{tabular}{r|r@{~}l}
Project & \% & wins for EMBLEM\\[0.1cm]

AMBER & 33 &   \RULEE{67pt} \\ 

HOOMD & 60 &  \RULEE{120pt} \\ 

RMG-PY  & 60 &  \RULEE{120pt}  \\ 

\cellcolor{gray!30}   SCIRUN  & 63 &   \RULEE{125pt}  \\ 

ABINIT & 63 &   \RULEE{125pt}  \\ 

\cellcolor{gray!30}  OMPI &  66 &   \RULEE{130pt}  \\ 

LIBMESH & 72 &  \RULEE{140pt}    \\  

MDANALYSIS & 72 &  \RULEE{140pt}   \\ 

LAMMPS & 75 &  \RULEE{150pt}  \\

\cellcolor{gray!30}   PSI4   & 80 &   \RULEE{160pt}  \\

XENON & 83 &\RULEE{170pt}

\end{tabular}
\end{table}

\noindent \textit{\underline{Notes:}} Research scientists typically do not produce documentation for the software they implement \cite{segal07_enduser, sanders08_risk}.
Further, there is a high personnel turnover rates in scientific software development \cite{carver06_hpc, segal07_problem}. As a result, there is a concern that CS software is harder to maintain. 

\noindent \textit{\underline{Modeling Assumptions:}} 
When code is hard to maintain,
developers who do work less frequently with the code are more prone to introduce defects
(rationale: the greater the complexity of the code, the greater the effort required to understand it).
Hence, one measure of the complexity of understanding code
is the difference in defect rates between core developers, also known as ``heroes''~\cite{agrawal2018we, goeminne2011evidence, torres2011analysis, robles2009evolution}, and everyone else.
Heroes are that  20\% group of the developers who usually make 80\% (or more) of the code changes

(Aside:   Majumder et al.~\cite{majumder19_heroes} found that such heroes are very common in open source projects. Their threshold for ``hero-ness'' are the 20\% of developers
who make 80\% of the changes.).

If the defect rate is much higher for non-heroes, that would
indicate that the code is so complex that it can only
be safely changed by those who have studied it in great detail.

\noindent \textit{\underline{Prediction: }} If CS code is hard to understand than SE code, we would expect that non-hero CS programmers would
introduce {\em more} defects into the software than non-hero SE programmers.

\noindent \textit{\underline{Observed:}} Majumder et al. \cite{majumder19_heroes} checked the heroes projects for both heroes and non-heroes contribution of defects within software development. They found that non-heroes introduced 30 to 90\% more defects per commit (25th-75th percentiles) in SE projects. Those
results can be see Figure \ref{fig:heroes} (left-hand-side).

We repeated their study for our 59 CS projects.  Figure \ref{fig:heroes} (right-hand-side)
shows those results. Each x-axis of that figure is one project so the hero and non-hero defect introduction rate (defects per commit) is the ratio of the blue to red numbers
at any particular x-value.
In those results, we observe that:
\bi
\item In CS projects, only 2/59 CS projects do non-heroes always introduce new defects with each commit (almost 1/3 for SE projects).
\item In SE projects, non-heroes's commits are far more likely (30\%-90\% for 25th-75th percentiles) than heroes to introduce new defects.
\item In CS projects, commits by non-heroes introduce new defects at nearly the same ratio as heroes (actually, 2\%-6\% less than for 25th-75th percentiles).
\ei
 
CS non-heroes introduce defects at much lower probability than in SE projects. Hence, we say:

\begin{RQ}
\textit{\underline{Conclusion:}} Measured in terms
of a number of defects introduced by each new commit, we \textbf{doubt} that the shared understanding of ``code'' is more difficult within  CS projects than SE projects.
\end{RQ}

% \begin{table}
% \small
% \begin{center}
% \caption{Percentiles seen in Figure \ref{fig:heroes}.}
% \label{tbl:heroes}
% \resizebox{1\linewidth}{!}{
% \begin{tabular}{c|c|r@{~}|r@{~}|r@{~}|r@{~}|r@{~}|r@{~}}
% & & \multicolumn{6}{c}{\textbf{Category}} \\
% \cline{3-8}
% &  & \multicolumn{3}{c|}{\textbf{SE Projects}} & \multicolumn{3}{c}{\textbf{CS Projects}}\\
% \cline{3-8}
% \textbf{Metric} & \begin{tabular}[c]{@{}c@{}} \textbf{Percentile} \end{tabular} & \begin{tabular}[c]{@{}c@{}} \textbf{Hero}\end{tabular} & \textbf{Non-Hero} & \begin{tabular}[c]{@{}c@{}} \textbf{Ratio}\end{tabular} & \textbf{Hero} & \textbf{Non-Hero} & \textbf{Ratio} \\ \hline

% \multirow{3}{*}{\begin{tabular}[l]{c} \rotatebox[origin=c]{90}{\parbox[c]{1.5cm}{\centering code interaction}} \end{tabular}}  & 
% 25th & 52 & 67 & 1.3 & 46 & 50 & 1.09  \\ [3pt]
% & 50th & 58 & 75 & 1.3 & 52 & 52 & 1 \\ [3pt]
% & 75th & 53 & 100 & 1.9 & 60 & 60 & 1 \\ [4pt]
% \hline
% \multirow{3}{*}{\begin{tabular}[l]{c} \rotatebox[origin=c]{90}{\parbox[c]{1.5cm}{\centering social interaction}} \end{tabular}}  & 
% 25th & 52 & 67 & 1.3 & 39 & 36 & 0.92  \\[3pt] 
% & 50th & 58 & 75 & 1.3 & 49 & 48 & 0.98 \\ [3pt]
% & 75th & 53 & 100 & 1.9 & 61 & 57 & 0.93 \\ [3pt]
% \end{tabular}}
% \end{center}
% \end{table}

\noindent \textit{\underline{Discussion:}} 
Cai et al. \cite{cai19_debt} argues that number of introduced bugs
per commit is {\em not} a measure of code comprehension.
In their case study, defect rates shot up after refactoring
precisely because (a)~developers now understood the code better so (b)~they were willing to make more changes so (c)~they
introduced more bugs. While their argument is certainly interesting, the Figure \ref{fig:heroes} (right-hand-side) results
are not a statement of defects {\em increased} after changes.
Rather, those results on defects ratios that are {\em the same} between two populations of programmers.

\subsection{Code Reuse} ~\\
\noindent \textit{\underline{Belief:}} 
Carver et al.~\cite{segal07_problem, carver06_hpc, Shull05_parallel, sanders08_risk} warn that there is little
code reuse in CS projects

\noindent \textit{\underline{Notes:}} 
Carver et al. report that scientific developers have a history of not adopting or re-use the software developed by others (or even their own). They 
say this is due to:

\bi
\item The structural assumptions from the others would be too strict and narrow \cite{carver06_hpc, basili08_hpc}
\item Most of the software is not built with comprehensibility requirement as the top priorities \cite{segal07_problem}. Hence, adapting old code for new domains is difficult.
\item
CS scientists believe that their time and efforts can be more conserved by being spent on implementing the new libraries and framework rather than understanding existing frameworks
\ei
\noindent ~\textit{\underline{Modeling Assumptions:}} To
measure reuse, we canmeasure the  code called
via libraries/ packages that come from outside of a repository. This is to say that the amount of external imports (EI) and files that have external imports (FEI) are indications of the reuse activities within the software. There are four attributes for this that we define below. For all of them, the higher the value the better reuse within their projects: 

\bi
\item \textit{IF\_Ratio} = EI / Total\_\#\_of\_Files
\item \textit{ILOC\_Ratio} = EI / LOC (total number lines of code)
\item \textit{II\_Ratio} = EI / Total\_\#\_of\_Imports 
\item \textit{FF\_Ratio} = FEI / Total\_\#\_of\_Files
\ei

\noindent~\textit{\underline{Prediction: }}CS projects have
less reuse than SE projects if the above ratios
are lower to CS than SE. 

\begin{table}[!t]
\small

\caption{Median and interquartile range (IQR) summary for four attributes portraying the reuse state of CS and SE projects.}
\label{tbl:reuse}
%\resizebox{1\linewidth}{!}{
\begin{tabular}{c|c|c|c}
\textbf{Metric} & \begin{tabular}[c]{@{}c@{}} \textbf{Project} \end{tabular} & \textbf{Median} & \textbf{IQR} \\ \hline
\multirow{2}{*}{\begin{tabular}[l]{c} %\rotatebox[origin=c]{90}{\parbox[c]{0.7cm}{\centering IF Ratio}} 
IF Ratio\end{tabular}} & 
CS & 3.2 & 1.6 \\ [2pt]
& SE & 2.9 & 1.6 \\ [2pt]
\hline
\multirow{2}{*}{\begin{tabular}[l]{c} %\rotatebox[origin=c]{90}{\parbox[c]{0.8cm}{\centering ILOC Ratio}}
ILOC Ratio
\end{tabular}} & 
SE & 13\textperthousand & 9\textperthousand \\ [2pt]
& CS & 10\textperthousand & 8\textperthousand \\ [3pt]
\hline
\multirow{2}{*}{\begin{tabular}[l]{c} %\rotatebox[origin=c]{90}{\parbox[c]{0.8cm}{\centering FF Ratio}}
FF Ratio\end{tabular}} & 
CS & 86\% & 19\% \\ [2pt]
& SE & 81\% & 19\% \\ [2pt]
\hline
\multirow{2}{*}{\begin{tabular}[l]{c} %\rotatebox[origin=c]{90}{\parbox[c]{0.7cm}{\centering II Ratio}} 
II Ratio\end{tabular}} & 
SE & 70\% & 27\%  \rule{0pt}{2.5ex} \\ [1.5pt]
& CS & 55\% & 19\% \\ [1.5pt]
\end{tabular}%}

\end{table}

\noindent \textit{\underline{Observed:}} Table \ref{tbl:reuse} summarizes the median and interquartile range for both CS and SE projects. The lines of code reuse is low for CS projects (just 10\%) but its nearly the same as SE projects (13\%). In fact,
after applying a Scott-Knott test\footnote{
Scott-Knott recursively divides treatments, stopping if
a significance test or an effect size test reports that sub-divisions are
not statistically different~\cite{mittas2013ranking, ghotra15}.
We use a bootstrap procedure to test for significance differences (at the 95\% confidence level) 
and the $\mathit{A12}$ test to check for small effects ($\mathit{A12} \ge 0.6$).
This procedure was selected
since it has been endorsed in the recent SE literature~\cite{mittas2013ranking,arcuri2011practical}.}, 
we can report that the SE projects are statistically
indistinguishable from CS projects, on all the metrics of Table \ref{tbl:reuse}. 
That, in this sample, we found no difference in the reuse rates
of SE and CS code. Hence: 
\begin{RQ} 
\textit{\underline{Conclusion:}} We \textbf{doubt} that CS reuses less code  than SE. 
\end{RQ}

\noindent \textit{\underline{Discussion:}} 
The ratios used here only reflect on code reuse.
Other kinds of reuse include design or conceptual reuse. 
Also missed by the above ratios is non-verbatim reuse (where code is reused, but modified).
Further, the above ratios may miss certain important code measures
(e.g. text-based, token-based, tree-based, metric-based, semantic and hybrid).

We did not explore those additional measures of reuse since
their implementation leads to $O(n^{(m-1)})$ complexity with $n$ as the current section of codes within the project and $m$ is the number of the projects to compare to. We hence leave reuse measurement in CS to  future work.

\subsection{Low Perceived Value} ~\\
\noindent \textit{\underline{Belief:}} 
Easterbrook et al. comment that even though CS codes
may be maintained for many years,
they are not perceived to have value within their own community~\cite{faulk09_secs, segal07_enduser, easterbrook_cs, boyle09_lessons}.

\noindent \textit{\underline{Notes:}} The social structures of the computational science community
typically reward new conclusions about physical phenomena much more than
details about the software used
to make those conclusions. This raises the concern since, as said in the introduction, the software is just as important a tool for modern science as, say, the test tube. If CS works need software,
but they perceive no value in that software, then the software may well be built and maintain in a sub-optimum manner~\cite{sanders08_risk}. 

\noindent \textit{\underline{Modeling Assumptions:}} 
One measure of software perceived value is its associated popularity within Github. This can be measured in many ways such as ratio of open to closed issues, or
numbers of stars or watchers or tags or forks. By consider the arrival rate of these measures with respect to the duration variable, and comparing those numbers between CS and SE, we can comment on how actively popular is a CS project compared to SE.

\noindent \textit{\underline{Prediction: }} 
According to this belief, CS projects should not be so actively popular as SE projects.

\noindent \textit{\underline{Observed:}} 
The Figure \ref{fig:comparison} showed a
level of activity for CS projects that rivals that of SE.
Except for duration, most of the indicators are similar or larger for CS than SE
(recall that even when the median CS results were lower, statistical tests showed that those differences
were not significantly distinguishable). Note that several of these indicators could be seen
to measure the popularity of a project. For example,
there are more closed releases that open issues which mean someone cares enough to work those issues.
Overall we can say:

\begin{RQ}
\textit{\underline{Conclusion:}} 
We  \textbf{doubt} that CS software is perceived by its community as having less value,
as compared to standard SE software.
\end{RQ}

\noindent \textit{\underline{Discussion:}} 
%Scientific software is developed by mostly Ph.D. students and postdocs that have high personnel turnover rates \cite{johan18_secs}. Moreover, due to grant-based funding schemes and research natures, the long-term vision is discouraged with ``quick and dirty'' solutions are more likely to be favored \cite{boyle09_lessons} which leads to short-lived software. 
% The result for the value aspect may surprise some folks. However, a trend in software-based research for both scientific developers or professional end-user is following the development of the projects and ultimately use such a project to curate necessary data, baseline results, and integrated framework that are relevant to their studies. Hence, the signal is clear by observing the similar distribution of \textit{Stars}, \textit{Forks}, and \textit{Watchers} (i.e. for following) and the greater distribution of \textit{Open \& Closed Issues} (i.e. for using) from CS's metrics to SE's metrics. 
% \noindent \textit{\underline{Threats of Validity:}} 
One threat to the validity of the above conclusion is that all of our sample of CS projects come from Github projects.
 There exist older existing systems and commercial projects that are not housed on Github. It is possible that those other systems are less popular than standard SE software. This would be an interesting area for future research.

%V
%comparison

% \noindent \textit{\underline{Threats of Validity:}} 

\subsection{Limited SE Training} ~\\
\noindent \textit{\underline{Belief:}} According to Segal et al.,
and others\cite{segal07_enduser, basili08_hpc, carver13_perception, easterbrook_cs, sanders08_risk}, few CS scientists are trained in SE.
This is a concern since that lack of training might lead to sub-optimum software
development practices.

\noindent \textit{\underline{Notes:}} 
The people who write the CS code usually
receive their degrees in
astronomy, astrophysics, chemistry, economics, genomics, molecular biology, oceanography, physics, political science, and many engineering fields.
That is, the primary field of study for these developers is {\em not}
software engineering. For many of these people,
learning SE is perceived as an excessive additional burden\cite{boyle09_lessons}. 

\noindent \textit{\underline{Modeling Assumptions:}} 
Successful training in SE is indicated by
\bi
\item We say an {\em efficient} software process is one that allows  people to work together, faster. 
\item The general quality of the software; e.g. the number of projects that pass the sanity checks of Table~\ref{tbl:sanity}.
\item The adoption of SE practices (e.g. an incremental development styles) can be inferred by comparing the distributions of different software development metrics between SE and CS.
\ei

\noindent \textit{\underline{Prediction: }} If this belief is valid, then more CS
projects should be poorly managed. Consequently, they would be less efficient. Also, fewer of them should
pass the sanity checks of Table~\ref{tbl:sanity}.
Further, we would not be able to detect current SE  practices within the CS project Github data.

\noindent \textit{\underline{Observed:}}
Recalling  the discussion about 
Figure~\ref{fig:comparison},
the case was made about if \S\ref{tion:data} that the CS development community
seems more {\em efficient} (as defiend above) than SE 

As to the sanity checks, 
two samples were used:
\bi
\item
We applied the sanity checks
of Table~\ref{tbl:sanity} to the 678 CS projects from~\S\ref{tion:data}. This selected 59 CS projects.
\item
Also, we took 50,000 SE Github projects (selected at random) and applied
the same sanity checks. This selected 1,300 projects.
\ei
This means that CS projects are over three times more likely to be {\em sane}: 

\centerline{\scalebox{1.1}{$\frac{\mathit{CS\_post\_pre\_sanity\_rate}}{\mathit{SE\_post\_pre\_sanity\_rate}} = \frac{\frac{\mathit{CS\_post\_sanity}}{\mathit{CS\_pre\_sanity}}}{\frac{\mathit{SE\_post\_sanity}}{\mathit{SE\_pre\_sanity}}} = $} \scalebox{1.5}{$\frac{\frac{59}{678}}{\frac{1,300}{50,000}} = $} 3.35 }

\begin{figure}
  \centering
  \includegraphics[width=\linewidth]{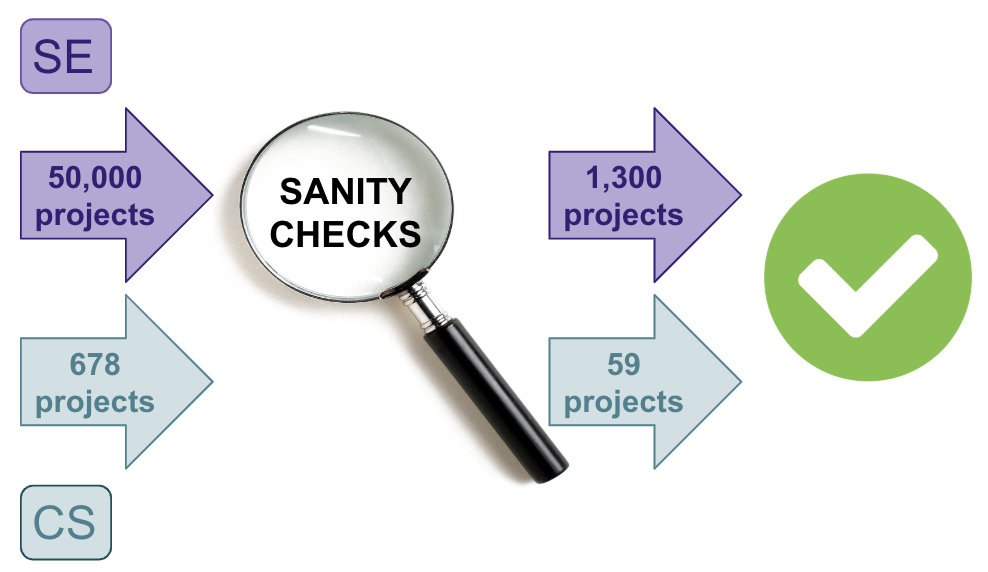} 
  \caption{Count of the numbers of SE and CS projects  before and after sanity checks.}
  \label{fig:sanity}
\end{figure}
 Also, recalling Figure \ref{fig:belief1} CS developers are observed to have a near-constant growth rate in their number of enhancements across their entire lifecycle. This observation is consistent with CS developers using
  contemporary continuous agile practices. 

From these observations, the picture that emerges is that CS developers are very good at adopting contemporary  SE approaches to SE development. More specifically, CS developers
use software engineering best practices   at least as much (and perhaps even more) that SE developers. 

\
Hence, we say: 
\begin{RQ}
\textit{\underline{Conclusion:}} We \textbf{doubt} this a lack of formal training in SE is inhibiting CS development
\end{RQ}

% \noindent \textit{\underline{Discussion:}} The view of training in SE for us is not solely about data structures and algorithm understandings (i.e. may not be even applicable when entering SE career in Sillicon Valley). It is about the real-world experience and intuitions that scientists can also pick up from working with colleagues (e.g. 1.a belief: requirements are not known upfront, agile philosophy). Similar to \S4.1's \textit{Discussion}, CS developers are well aware that SE skills are appealing to high-profile companies in the industry.  

% Moreover, we also offer additional evidence to doubt this belief in \S5.2 and \S5.3. CS novices and outsiders understand the source code than SE ones to contribute significantly less defects. CS projects reuse as much code as SE projects.   

%\subsection{External Validity}

%Methodology to understand the difference between software development domains in SE. Not our final conclusions or definitions, all are threatened by external validities but our conclusions are reproducible and our analysis can be repeated when new data arrives. 

\section{Discussion}
This section reflects on the 13 beliefs study means for
applying SE methods to CS.

Firstly, there is much SE that can be applied to CS. We saw many times in this study that CS developers are very interested and aware of SE methods (e.g. agile philosophy in \S4.1 and modern techniques in \S5.1). Computational Science is a rich domain within which SE tools can be very useful.

That said,  we offer one word of caution about moving SE's tools and methods to CS. The discussion in \S\ref{terms} warned that sometimes basic terminology can be different in SE to CS. It is, therefore, wise to spend some time checking domain terminology. The incremental data mining tool described in \S\ref{terms} is one way to reduce the time and cost
involved in performing such checks. More generally, this calls for attention to not apply off-the-shelf method when moving to a different community, it is more useful to tailor SE methods for the CS community.

Secondly, contrary to much prior pessimism, the overall message of this paper is that CS software development is at least as successful as standard development practices seen in SE projects (e.g. code understanding in \S6.2, perceived value in \S6.4, and overall SE background in \S6.5). This means that while CS can take useful tools and insights for SE, there is also room for insights and tools to flow backward from CS to help SE.
In particular, the relatively lower defect introduction rates seen in Figure~\ref{fig:heroes} are worthy of further study. Perhaps there is something SE can learn from CS about how to design systems that are less buggy.

Thirdly, recalling the discussion about lack of requirements in \S\ref{rments}, it would appear that better methods for requirements engineering may not the most cost-effective thing that SE can offer CS. To be sure, in some CS domains such as hydrology (where CS developers work closely with civil engineers), there is space for better requirements engineering. But overall, \S\ref{rments} is saying that if there is only {\em one} thing you try to improve, changes to requirements engineering many not yield the most benefits for CS.

As to other parts of the development lifecycle,  recalling the discussion about verification and validation in \S\ref{vv}, CS would most benefit from a different kind of testing device. Standard SE is to divide testing into the unit and system testing. \S\ref{vv} says there is a third layer of testing that we might call science testing. CS debug tools need to be augmented with (e.g.)  physical knowledge that can detect violations of physical properties.

Apart from requirements and testing, another major part of the software lifecycle is development and operations. These are two areas that seem to offer the most benefit for new research. For example, many CS projects are ``glue'' codes that allow other people to run their experimental application code on some complex platform (with one of the benefit being accessible to other tools within the research community). When that code crashes, it is a triage problem to decide which team needs to fix the code (the ``glue'' developers or the application developers). This is one example of the kind of operational support that would be beneficial to explore. Clearly, there are many more possibilities in this exciting area. In fact, the three beliefs that we did not get to investigate, mentioned in \S2.4, in this study will be our future work in action right away.

\section{Conclusion}

%he premise of the prior research \cite{johan18_secs} is underlying shortcomings of existing approaches for bridging the gap between Software Engineering and Computational Science can be identified by the 13 recurring characteristics or beliefs.
Through a quantitative investigation on 59 projects, we have found several disconnects between current data
and some-held beliefs about computational science.
Why are so many of those older beliefs not supportable?
We argue that  the  nature of the CS software development is changing. 
For example, contrary to much prior pessimism,
CS developers are now very
aware of SE methods. 
We can see much evidence that CS developers are making
extensive use of SE methods.

 The current work here lays out highlighted perspectives, quantitative evidence to clarify existing beliefs about scientific software development.
 We hope these results  prompt a fresh examination of the nature of SE in  CS which, in turn, might suggests 
  new   specialized supporting tools for CS.
For example, requirements and unit and system testing  are considered hot topics in in the SE community.
But for CS projects,   studying  (a)~scientific testing (b)~development and
(c)~operations might be comparatively more useful.

% Therefore, more
% research on this topic is needed, especially to empirically evaluate the 
% gains in productivity and credibility achieved for scientific software by such
% SE approaches. 

\section{Acknowledgments}

We thank the   CS community
from the Molecular Sciences Software Institute (MOLSSI), and the Science Gateways Community Institute (SGCI)) for
their assistance with this work.

This work was partially funded by 
%an NSF CISE Grant \#1826574 and \#1931425.
blinded for review.

\balance
\bibliographystyle{ACM-Reference-Format}
\bibliography{sample.bbl}


\begin{thebibliography}{49}

%%% the Web address.

\ifx \showCODEN    \undefined \def \showCODEN     #1{\unskip}     \fi
\ifx \showDOI      \undefined \def \showDOI       #1{#1}\fi
\ifx \showISBNx    \undefined \def \showISBNx     #1{\unskip}     \fi
\ifx \showISBNxiii \undefined \def \showISBNxiii  #1{\unskip}     \fi
\ifx \showISSN     \undefined \def \showISSN      #1{\unskip}     \fi
\ifx \showLCCN     \undefined \def \showLCCN      #1{\unskip}     \fi
\ifx \shownote     \undefined \def \shownote      #1{#1}          \fi
\ifx \showarticletitle \undefined \def \showarticletitle #1{#1}   \fi
\ifx \showURL      \undefined \def \showURL       {\relax}        \fi
% invisible to TeX
\providecommand\bibfield[2]{#2}
\providecommand\bibinfo[2]{#2}
\providecommand\natexlab[1]{#1}
\providecommand\showeprint[2][]{arXiv:#2}

\bibitem[\protect\citeauthoryear{Agrawal, Rahman, Krishna, Sobran, and
  Menzies}{Agrawal et~al\mbox{.}}{2018}]%
        {agrawal2018we}
\bibfield{author}{\bibinfo{person}{A. Agrawal}, \bibinfo{person}{A. Rahman},
  \bibinfo{person}{R. Krishna}, \bibinfo{person}{A. Sobran}, {and}
  \bibinfo{person}{T. Menzies}.} \bibinfo{year}{2018}\natexlab{}.
\newblock \showarticletitle{We don't need another hero?: the impact of heroes
  on software development}. In \bibinfo{booktitle}{\emph{ICSE}}.
\newblock


\bibitem[\protect\citeauthoryear{Arcuri and Briand}{Arcuri and Briand}{2011}]%
        {arcuri2011practical}
\bibfield{author}{\bibinfo{person}{A. Arcuri} {and} \bibinfo{person}{L.
  Briand}.} \bibinfo{year}{2011}\natexlab{}.
\newblock \showarticletitle{A practical guide for using statistical tests to
  assess randomized algorithms in software engineering}. In
  \bibinfo{booktitle}{\emph{ICSE}}. IEEE.
\newblock


\bibitem[\protect\citeauthoryear{Basili, Carver, Cruzes, Hochstein,
  Hollingsworth, Shull, and Zelkowitz}{Basili et~al\mbox{.}}{2008}]%
        {basili08_hpc}
\bibfield{author}{\bibinfo{person}{V.~R. Basili}, \bibinfo{person}{J.~C.
  Carver}, \bibinfo{person}{D. Cruzes}, \bibinfo{person}{L.~M. Hochstein},
  \bibinfo{person}{J.~K. Hollingsworth}, \bibinfo{person}{F. Shull}, {and}
  \bibinfo{person}{M.~V. Zelkowitz}.} \bibinfo{year}{2008}\natexlab{}.
\newblock \showarticletitle{Understanding the High-Performance-Computing
  Community: A Software Engineer's Perspective}.
\newblock \bibinfo{journal}{\emph{IEEE Software}} (\bibinfo{year}{2008}).
\newblock


\bibitem[\protect\citeauthoryear{{Bird}, {Rigby}, {Barr}, {Hamilton}, {German},
  and {Devanbu}}{{Bird} et~al\mbox{.}}{2009}]%
        {bird09promise}
\bibfield{author}{\bibinfo{person}{C. {Bird}}, \bibinfo{person}{P.~C. {Rigby}},
  \bibinfo{person}{E.~T. {Barr}}, \bibinfo{person}{D.~J. {Hamilton}},
  \bibinfo{person}{D.~M. {German}}, {and} \bibinfo{person}{P. {Devanbu}}.}
  \bibinfo{year}{2009}\natexlab{}.
\newblock \showarticletitle{The promises and perils of mining git}. In
  \bibinfo{booktitle}{\emph{Mining Software Repositories}}.
\newblock


\bibitem[\protect\citeauthoryear{Brooks~Jr}{Brooks~Jr}{1995}]%
        {brooks1995mythical}
\bibfield{author}{\bibinfo{person}{Frederick~P Brooks~Jr}.}
  \bibinfo{year}{1995}\natexlab{}.
\newblock \showarticletitle{The mythical man-month (anniversary ed.)}.
\newblock  (\bibinfo{year}{1995}).
\newblock


\bibitem[\protect\citeauthoryear{Carver, Heaton, Hochstein, and
  Bartlett}{Carver et~al\mbox{.}}{2013}]%
        {carver13_perception}
\bibfield{author}{\bibinfo{person}{J. Carver}, \bibinfo{person}{D. Heaton},
  \bibinfo{person}{L. Hochstein}, {and} \bibinfo{person}{R. Bartlett}.}
  \bibinfo{year}{2013}\natexlab{}.
\newblock \showarticletitle{Self-Perceptions about Software Engineering: A
  Survey of Scientists and Engineers}.
\newblock \bibinfo{journal}{\emph{Computing in Science Engineering}}
  (\bibinfo{year}{2013}).
\newblock


\bibitem[\protect\citeauthoryear{Carver, Hochstein, Kendall, Nakamura,
  Zelkowitz, R~Basili, and Post}{Carver et~al\mbox{.}}{2006}]%
        {carver06_hpc}
\bibfield{author}{\bibinfo{person}{Jeff Carver}, \bibinfo{person}{Lorin
  Hochstein}, \bibinfo{person}{Richard Kendall}, \bibinfo{person}{Taiga
  Nakamura}, \bibinfo{person}{Marvin Zelkowitz}, \bibinfo{person}{Victor
  R~Basili}, {and} \bibinfo{person}{Douglass Post}.}
  \bibinfo{year}{2006}\natexlab{}.
\newblock \showarticletitle{Observations about software development for high
  end computing}.
\newblock \bibinfo{journal}{\emph{CT Watch Quarterly}} (\bibinfo{year}{2006}).
\newblock


\bibitem[\protect\citeauthoryear{Carver, Kendall, Squires, and Post}{Carver
  et~al\mbox{.}}{2007}]%
        {carver07_environment}
\bibfield{author}{\bibinfo{person}{J.~C. Carver}, \bibinfo{person}{R.~P.
  Kendall}, \bibinfo{person}{S.~E. Squires}, {and} \bibinfo{person}{D.~E.
  Post}.} \bibinfo{year}{2007}\natexlab{}.
\newblock \showarticletitle{Software Development Environments for Scientific
  and Engineering Software: A Series of Case Studies}. In
  \bibinfo{booktitle}{\emph{29th International Conference on Software
  Engineering (ICSE'07)}}. \bibinfo{pages}{550--559}.
\newblock


\bibitem[\protect\citeauthoryear{Devanbu, Zimmermann, and Bird}{Devanbu
  et~al\mbox{.}}{2016}]%
        {dev16}
\bibfield{author}{\bibinfo{person}{Prem Devanbu}, \bibinfo{person}{Thomas
  Zimmermann}, {and} \bibinfo{person}{Christian Bird}.}
  \bibinfo{year}{2016}\natexlab{}.
\newblock \showarticletitle{Belief \& Evidence in Empirical Software
  Engineering}. In \bibinfo{booktitle}{\emph{Proceedings of the 38th
  International Conference on Software Engineering}}
  \emph{(\bibinfo{series}{ICSE ’16})}. \bibinfo{publisher}{Association for
  Computing Machinery}, \bibinfo{address}{New York, NY, USA},
  \bibinfo{pages}{108–119}.
\newblock
\showISBNx{9781450339001}
\urldef\tempurl%
\url{https://doi.org/10.1145/2884781.2884812}
\showDOI{\tempurl}


\bibitem[\protect\citeauthoryear{{Easterbrook} and {Johns}}{{Easterbrook} and
  {Johns}}{2009}]%
        {easterbrook_cs}
\bibfield{author}{\bibinfo{person}{S.~M. {Easterbrook}} {and}
  \bibinfo{person}{T.~C. {Johns}}.} \bibinfo{year}{2009}\natexlab{}.
\newblock \showarticletitle{Engineering the Software for Understanding Climate
  Change}.
\newblock \bibinfo{journal}{\emph{Computing in Science Engineering}}
  (\bibinfo{year}{2009}).
\newblock


\bibitem[\protect\citeauthoryear{Efron and Tibshirani}{Efron and
  Tibshirani}{1994}]%
        {efron94}
\bibfield{author}{\bibinfo{person}{Bradley Efron} {and}
  \bibinfo{person}{Robert~J Tibshirani}.} \bibinfo{year}{1994}\natexlab{}.
\newblock \bibinfo{booktitle}{\emph{An introduction to the bootstrap}}.
\newblock \bibinfo{address}{London}.
\newblock


\bibitem[\protect\citeauthoryear{Ghotra, McIntosh, and Hassan}{Ghotra
  et~al\mbox{.}}{[n. d.]}]%
        {ghotra15}
\bibfield{author}{\bibinfo{person}{B. Ghotra}, \bibinfo{person}{S. McIntosh},
  {and} \bibinfo{person}{A.~E. Hassan}.} \bibinfo{year}{[n. d.]}\natexlab{}.
\newblock \showarticletitle{Revisiting the Impact of Classification Techniques
  on the Performance of Defect Prediction Models}. In
  \bibinfo{booktitle}{\emph{2015 37th ICSE}}.
\newblock


\bibitem[\protect\citeauthoryear{Goeminne and Mens}{Goeminne and Mens}{2011}]%
        {goeminne2011evidence}
\bibfield{author}{\bibinfo{person}{Mathieu Goeminne} {and} \bibinfo{person}{Tom
  Mens}.} \bibinfo{year}{2011}\natexlab{}.
\newblock \showarticletitle{Evidence for the pareto principle in open source
  software activity}. In \bibinfo{booktitle}{\emph{the Joint Porceedings of the
  1st International workshop on Model Driven Software Maintenance and 5th
  International Workshop on Software Quality and Maintainability}}.
  \bibinfo{pages}{74--82}.
\newblock


\bibitem[\protect\citeauthoryear{Heaton and Carver}{Heaton and Carver}{2015}]%
        {heaton15_lit}
\bibfield{author}{\bibinfo{person}{Dustin Heaton} {and}
  \bibinfo{person}{Jeffrey~C. Carver}.} \bibinfo{year}{2015}\natexlab{}.
\newblock \showarticletitle{Claims about the use of software engineering
  practices in science: A systematic literature review}.
\newblock \bibinfo{journal}{\emph{Information and Software Technology}}
  \bibinfo{volume}{67} (\bibinfo{year}{2015}), \bibinfo{pages}{207 -- 219}.
\newblock


\bibitem[\protect\citeauthoryear{Heroux, Bartlett, Howle, Hoekstra, Hu, Kolda,
  Lehoucq, Long, Pawlowski, Phipps, Salinger, Thornquist, Tuminaro,
  Willenbring, Williams, and Stanley}{Heroux et~al\mbox{.}}{2005}]%
        {kendall05_C}
\bibfield{author}{\bibinfo{person}{Michael Heroux}, \bibinfo{person}{Roscoe
  Bartlett}, \bibinfo{person}{Victoria Howle}, \bibinfo{person}{Robert
  Hoekstra}, \bibinfo{person}{Jonathan Hu}, \bibinfo{person}{Tamara Kolda},
  \bibinfo{person}{Richard Lehoucq}, \bibinfo{person}{Katharine Long},
  \bibinfo{person}{Roger Pawlowski}, \bibinfo{person}{Eric Phipps},
  \bibinfo{person}{Andrew Salinger}, \bibinfo{person}{Heidi Thornquist},
  \bibinfo{person}{R. Tuminaro}, \bibinfo{person}{James Willenbring},
  \bibinfo{person}{Alan Williams}, {and} \bibinfo{person}{Kendall Stanley}.}
  \bibinfo{year}{2005}\natexlab{}.
\newblock \showarticletitle{An overview of the Trilinos Project}.
\newblock \bibinfo{journal}{\emph{ACM Trans. Math. Softw.}}
  \bibinfo{volume}{31} (\bibinfo{date}{09} \bibinfo{year}{2005}),
  \bibinfo{pages}{397--423}.
\newblock
\urldef\tempurl%
\url{https://doi.org/10.1145/1089014.1089021}
\showDOI{\tempurl}


\bibitem[\protect\citeauthoryear{Hindle, German, and Holt}{Hindle
  et~al\mbox{.}}{2008}]%
        {hindle08_largecommits}
\bibfield{author}{\bibinfo{person}{A. Hindle}, \bibinfo{person}{D.~M. German},
  {and} \bibinfo{person}{R. Holt}.} \bibinfo{year}{2008}\natexlab{}.
\newblock \showarticletitle{What Do Large Commits Tell Us?: A Taxonomical Study
  of Large Commits} \emph{(\bibinfo{series}{MSR})}.
\newblock


\bibitem[\protect\citeauthoryear{{Hochstein}, {Carver}, {Shull}, {Asgari},
  {Basili}, {Hollingsworth}, and {Zelkowitz}}{{Hochstein}
  et~al\mbox{.}}{2005}]%
        {Shull05_parallel}
\bibfield{author}{\bibinfo{person}{L. {Hochstein}}, \bibinfo{person}{J.
  {Carver}}, \bibinfo{person}{F. {Shull}}, \bibinfo{person}{S. {Asgari}},
  \bibinfo{person}{V. {Basili}}, \bibinfo{person}{J.~K. {Hollingsworth}}, {and}
  \bibinfo{person}{M.~V. {Zelkowitz}}.} \bibinfo{year}{2005}\natexlab{}.
\newblock \showarticletitle{Parallel Programmer Productivity: A Case Study of
  Novice Parallel Programmers}. In \bibinfo{booktitle}{\emph{SC '05:
  Proceedings of the 2005 ACM/IEEE Conference on Supercomputing}}.
  \bibinfo{pages}{35--35}.
\newblock
\showISSN{null}
\urldef\tempurl%
\url{https://doi.org/10.1109/SC.2005.53}
\showDOI{\tempurl}


\bibitem[\protect\citeauthoryear{Johanson and Hasselbring}{Johanson and
  Hasselbring}{2018}]%
        {johan18_secs}
\bibfield{author}{\bibinfo{person}{A. Johanson} {and} \bibinfo{person}{W.
  Hasselbring}.} \bibinfo{year}{2018}\natexlab{}.
\newblock \showarticletitle{Software Engineering for Computational Science:
  Past, Present, Future}.
\newblock \bibinfo{journal}{\emph{Computing in Science Engineering}}
  (\bibinfo{year}{2018}).
\newblock


\bibitem[\protect\citeauthoryear{Kalliamvakou, Gousios, Blincoe, Singer,
  German, and Damian}{Kalliamvakou et~al\mbox{.}}{2014}]%
        {Kalliamvakou:2014}
\bibfield{author}{\bibinfo{person}{E. Kalliamvakou}, \bibinfo{person}{G.
  Gousios}, \bibinfo{person}{K. Blincoe}, \bibinfo{person}{L. Singer},
  \bibinfo{person}{D. German}, {and} \bibinfo{person}{D. Damian}.}
  \bibinfo{year}{2014}\natexlab{}.
\newblock \showarticletitle{The Promises and Perils of Mining GitHub}. In
  \bibinfo{booktitle}{\emph{MSR}}.
\newblock


\bibitem[\protect\citeauthoryear{Kalliamvakou, Gousios, Blincoe, Singer,
  M~German, and Damian}{Kalliamvakou et~al\mbox{.}}{2015}]%
        {eirini15promise}
\bibfield{author}{\bibinfo{person}{E. Kalliamvakou}, \bibinfo{person}{G.
  Gousios}, \bibinfo{person}{K. Blincoe}, \bibinfo{person}{L. Singer},
  \bibinfo{person}{D. M~German}, {and} \bibinfo{person}{D. Damian}.}
  \bibinfo{year}{2015}\natexlab{}.
\newblock \showarticletitle{The Promises and Perils of Mining GitHub (Extended
  Version)}.
\newblock \bibinfo{journal}{\emph{EMSE}} (\bibinfo{year}{2015}).
\newblock


\bibitem[\protect\citeauthoryear{Kamei, Shihab, Adams, Hassan, Mockus, Sinha,
  and Ubayashi}{Kamei et~al\mbox{.}}{2013}]%
        {kamei12_jit}
\bibfield{author}{\bibinfo{person}{Y. Kamei}, \bibinfo{person}{E. Shihab},
  \bibinfo{person}{B. Adams}, \bibinfo{person}{A.~E. Hassan},
  \bibinfo{person}{A. Mockus}, \bibinfo{person}{A. Sinha}, {and}
  \bibinfo{person}{N. Ubayashi}.} \bibinfo{year}{2013}\natexlab{}.
\newblock \showarticletitle{A large-scale empirical study of just-in-time
  quality assurance}.
\newblock \bibinfo{journal}{\emph{TSE}} (\bibinfo{year}{2013}).
\newblock


\bibitem[\protect\citeauthoryear{Kanewala and Bieman}{Kanewala and
  Bieman}{2013}]%
        {kanewala13_testing}
\bibfield{author}{\bibinfo{person}{U. Kanewala} {and} \bibinfo{person}{J.~M.
  Bieman}.} \bibinfo{year}{2013}\natexlab{}.
\newblock \showarticletitle{Using machine learning techniques to detect
  metamorphic relations for programs without test oracles}. In
  \bibinfo{booktitle}{\emph{ISSRE}}.
\newblock


\bibitem[\protect\citeauthoryear{Kaplan and Norton}{Kaplan and Norton}{1996}]%
        {kaplan1996using}
\bibfield{author}{\bibinfo{person}{Robert~S Kaplan} {and}
  \bibinfo{person}{David~P Norton}.} \bibinfo{year}{1996}\natexlab{}.
\newblock \showarticletitle{Using the Balanced Scorecard as a Strategic
  Management System}.
\newblock \bibinfo{journal}{\emph{Harvard Business Review}}
  (\bibinfo{year}{1996}).
\newblock


\bibitem[\protect\citeauthoryear{{Killcoyne} and {Boyle}}{{Killcoyne} and
  {Boyle}}{2009}]%
        {boyle09_lessons}
\bibfield{author}{\bibinfo{person}{S. {Killcoyne}} {and} \bibinfo{person}{J.
  {Boyle}}.} \bibinfo{year}{2009}\natexlab{}.
\newblock \showarticletitle{Managing Chaos: Lessons Learned Developing Software
  in the Life Sciences}.
\newblock \bibinfo{journal}{\emph{Computing in Science Engineering}}
  \bibinfo{volume}{11}, \bibinfo{number}{6} (\bibinfo{date}{Nov}
  \bibinfo{year}{2009}), \bibinfo{pages}{20--29}.
\newblock
\showISSN{1558-366X}
\urldef\tempurl%
\url{https://doi.org/10.1109/MCSE.2009.198}
\showDOI{\tempurl}


\bibitem[\protect\citeauthoryear{Kim, Whitehead, and Zhang}{Kim
  et~al\mbox{.}}{2008}]%
        {Kim08changes}
\bibfield{author}{\bibinfo{person}{S. Kim}, \bibinfo{person}{E.~J. Whitehead,
  Jr.}, {and} \bibinfo{person}{Y. Zhang}.} \bibinfo{year}{2008}\natexlab{}.
\newblock \showarticletitle{Classifying Software Changes: Clean or Buggy?}
\newblock \bibinfo{journal}{\emph{TSE}} (\bibinfo{year}{2008}).
\newblock


\bibitem[\protect\citeauthoryear{M.~Shanley}{M.~Shanley}{2013}]%
        {nobel_2013}
\bibfield{author}{\bibinfo{person}{S.~Nordenstam M.~Shanley}.}
  \bibinfo{year}{2013}\natexlab{}.
\newblock \bibinfo{title}{Scientists who took chemistry into cyberspace win
  Nobel Prize}.
\newblock
\newblock


\bibitem[\protect\citeauthoryear{Majumder, Chakraborty, Agrawal, and
  Menzies}{Majumder et~al\mbox{.}}{2019a}]%
        {Majumder19}
\bibfield{author}{\bibinfo{person}{Suvodeep Majumder},
  \bibinfo{person}{Joymallya Chakraborty}, \bibinfo{person}{Amritanshu
  Agrawal}, {and} \bibinfo{person}{Tim Menzies}.}
  \bibinfo{year}{2019}\natexlab{a}.
\newblock \showarticletitle{Why Software Projects need Heroes (Lessons Learned
  from 1100+ Projects)}.
\newblock \bibinfo{journal}{\emph{CoRR}} (\bibinfo{year}{2019}).
\newblock


\bibitem[\protect\citeauthoryear{Majumder, Chakraborty, Agrawal, and
  Menzies}{Majumder et~al\mbox{.}}{2019b}]%
        {majumder19_heroes}
\bibfield{author}{\bibinfo{person}{Suvodeep Majumder},
  \bibinfo{person}{Joymallya Chakraborty}, \bibinfo{person}{Amritanshu
  Agrawal}, {and} \bibinfo{person}{Tim Menzies}.}
  \bibinfo{year}{2019}\natexlab{b}.
\newblock \showarticletitle{Why Software Projects need Heroes (Lessons Learned
  from 1100+ Projects)}.
\newblock \bibinfo{journal}{\emph{CoRR}}  \bibinfo{volume}{abs/1904.09954}
  (\bibinfo{year}{2019}).
\newblock
\showeprint[arxiv]{1904.09954}
\urldef\tempurl%
\url{http://arxiv.org/abs/1904.09954}
\showURL{%
\tempurl}


\bibitem[\protect\citeauthoryear{Menzies, Nichols, Shull, and Layman}{Menzies
  et~al\mbox{.}}{2017}]%
        {menzies17}
\bibfield{author}{\bibinfo{person}{Tim Menzies}, \bibinfo{person}{William
  Nichols}, \bibinfo{person}{Forrest Shull}, {and} \bibinfo{person}{Lucas
  Layman}.} \bibinfo{year}{2017}\natexlab{}.
\newblock \showarticletitle{Are Delayed Issues Harder to Resolve? Revisiting
  Cost-to-Fix of Defects throughout the Lifecycle}.
\newblock \bibinfo{journal}{\emph{Empirical Softw. Engg.}}
  \bibinfo{volume}{22}, \bibinfo{number}{4} (\bibinfo{date}{Aug.}
  \bibinfo{year}{2017}), \bibinfo{pages}{1903–1935}.
\newblock
\showISSN{1382-3256}
\urldef\tempurl%
\url{https://doi.org/10.1007/s10664-016-9469-x}
\showDOI{\tempurl}


\bibitem[\protect\citeauthoryear{Merali}{Merali}{2010}]%
        {merali10_error}
\bibfield{author}{\bibinfo{person}{Zeeya Merali}.}
  \bibinfo{year}{2010}\natexlab{}.
\newblock \showarticletitle{Computational science: Error, why scientific
  programming does not compute}.
\newblock \bibinfo{journal}{\emph{Nature}} \bibinfo{volume}{467},
  \bibinfo{number}{7317} (\bibinfo{year}{2010}).
\newblock
\urldef\tempurl%
\url{https://doi.org/10.1038/467775a}
\showDOI{\tempurl}


\bibitem[\protect\citeauthoryear{Mittas and Angelis}{Mittas and
  Angelis}{2013}]%
        {mittas2013ranking}
\bibfield{author}{\bibinfo{person}{N. Mittas} {and} \bibinfo{person}{L.
  Angelis}.} \bibinfo{year}{2013}\natexlab{}.
\newblock \showarticletitle{Ranking and clustering software cost estimation
  models through a multiple comparisons algorithm}.
\newblock \bibinfo{journal}{\emph{TSE}} (\bibinfo{year}{2013}).
\newblock


\bibitem[\protect\citeauthoryear{Mockus and Votta}{Mockus and Votta}{2000}]%
        {mockus00changeskeys}
\bibfield{author}{\bibinfo{person}{A. Mockus} {and} \bibinfo{person}{L.
  Votta}.} \bibinfo{year}{2000}\natexlab{}.
\newblock \showarticletitle{Identifying reasons for software changes using
  historic databases}. In \bibinfo{booktitle}{\emph{ICPC}}.
\newblock


\bibitem[\protect\citeauthoryear{Munaiah, Kroh, Cabrey, and Nagappan}{Munaiah
  et~al\mbox{.}}{2017}]%
        {munaiah17curating}
\bibfield{author}{\bibinfo{person}{N. Munaiah}, \bibinfo{person}{S. Kroh},
  \bibinfo{person}{C. Cabrey}, {and} \bibinfo{person}{M. Nagappan}.}
  \bibinfo{year}{2017}\natexlab{}.
\newblock \showarticletitle{Curating GitHub for Engineered Software Projects}.
\newblock \bibinfo{journal}{\emph{EMSE}} (\bibinfo{year}{2017}).
\newblock


\bibitem[\protect\citeauthoryear{{Nayebi}, {Cai}, {Kazman}, {Ruhe}, {Feng},
  {Carlson}, and {Chew}}{{Nayebi} et~al\mbox{.}}{2019}]%
        {cai19_debt}
\bibfield{author}{\bibinfo{person}{M. {Nayebi}}, \bibinfo{person}{Y. {Cai}},
  \bibinfo{person}{R. {Kazman}}, \bibinfo{person}{G. {Ruhe}},
  \bibinfo{person}{Q. {Feng}}, \bibinfo{person}{C. {Carlson}}, {and}
  \bibinfo{person}{F. {Chew}}.} \bibinfo{year}{2019}\natexlab{}.
\newblock \showarticletitle{A Longitudinal Study of Identifying and Paying Down
  Architecture Debt}.
\newblock  (\bibinfo{year}{2019}).
\newblock
\urldef\tempurl%
\url{https://doi.org/10.1109/ICSE-SEIP.2019.00026}
\showDOI{\tempurl}


\bibitem[\protect\citeauthoryear{Prabhu, Jablin, Raman, Zhang, Huang, Kim,
  Johnson, Liu, Ghosh, Beard, Oh, Zoufaly, Walker, and August}{Prabhu
  et~al\mbox{.}}{2011}]%
        {Prabhu11_cssurvey}
\bibfield{author}{\bibinfo{person}{Prakash Prabhu}, \bibinfo{person}{Thomas~B.
  Jablin}, \bibinfo{person}{Arun Raman}, \bibinfo{person}{Yun Zhang},
  \bibinfo{person}{Jialu Huang}, \bibinfo{person}{Hanjun Kim},
  \bibinfo{person}{Nick~P. Johnson}, \bibinfo{person}{Feng Liu},
  \bibinfo{person}{Soumyadeep Ghosh}, \bibinfo{person}{Stephen Beard},
  \bibinfo{person}{Taewook Oh}, \bibinfo{person}{Matthew Zoufaly},
  \bibinfo{person}{David Walker}, {and} \bibinfo{person}{David~I. August}.}
  \bibinfo{year}{2011}\natexlab{}.
\newblock \showarticletitle{A Survey of the Practice of Computational Science}.
  In \bibinfo{booktitle}{\emph{State of the Practice Reports}}.
  \bibinfo{publisher}{ACM}.
\newblock


\bibitem[\protect\citeauthoryear{{Ragan-Kelley}, {Perez}, {Granger}, {Kluyver},
  {Ivanov}, {Frederic}, and {Bussonnier}}{{Ragan-Kelley} et~al\mbox{.}}{2014}]%
        {ragan14_pythoncs}
\bibfield{author}{\bibinfo{person}{M. {Ragan-Kelley}}, \bibinfo{person}{F.
  {Perez}}, \bibinfo{person}{B. {Granger}}, \bibinfo{person}{T. {Kluyver}},
  \bibinfo{person}{P. {Ivanov}}, \bibinfo{person}{J. {Frederic}}, {and}
  \bibinfo{person}{M. {Bussonnier}}.} \bibinfo{year}{2014}\natexlab{}.
\newblock \showarticletitle{{The Jupyter/IPython architecture: a unified view
  of computational research, from interactive exploration to communication and
  publication.}}. In \bibinfo{booktitle}{\emph{AGU Fall Meeting Abstracts}}.
\newblock


\bibitem[\protect\citeauthoryear{Robles, Gonzalez-Barahona, and Herraiz}{Robles
  et~al\mbox{.}}{2009}]%
        {robles2009evolution}
\bibfield{author}{\bibinfo{person}{Gregorio Robles}, \bibinfo{person}{Jesus~M
  Gonzalez-Barahona}, {and} \bibinfo{person}{Israel Herraiz}.}
  \bibinfo{year}{2009}\natexlab{}.
\newblock \showarticletitle{Evolution of the core team of developers in libre
  software projects}. In \bibinfo{booktitle}{\emph{Mining Software
  Repositories, 2009. MSR'09. 6th IEEE International Working Conference on}}.
  IEEE, \bibinfo{pages}{167--170}.
\newblock


\bibitem[\protect\citeauthoryear{Sanders and Kelly}{Sanders and Kelly}{2008}]%
        {sanders08_risk}
\bibfield{author}{\bibinfo{person}{Rebecca Sanders} {and}
  \bibinfo{person}{Diane Kelly}.} \bibinfo{year}{2008}\natexlab{}.
\newblock \showarticletitle{Dealing with Risk in Scientific Software
  Development}.
\newblock \bibinfo{journal}{\emph{Software, IEEE}}  \bibinfo{volume}{25}
  (\bibinfo{date}{08} \bibinfo{year}{2008}), \bibinfo{pages}{21 -- 28}.
\newblock
\urldef\tempurl%
\url{https://doi.org/10.1109/MS.2008.84}
\showDOI{\tempurl}


\bibitem[\protect\citeauthoryear{Schouten, Shlomo, and Skinner}{Schouten
  et~al\mbox{.}}{2010}]%
        {schouten2010indicators}
\bibfield{author}{\bibinfo{person}{Barry Schouten}, \bibinfo{person}{Natalie
  Shlomo}, {and} \bibinfo{person}{Chris Skinner}.}
  \bibinfo{year}{2010}\natexlab{}.
\newblock \showarticletitle{Indicators for monitoring and improving
  representativeness of response}.
\newblock  (\bibinfo{year}{2010}).
\newblock


\bibitem[\protect\citeauthoryear{Segal}{Segal}{2005}]%
        {segal05_ss}
\bibfield{author}{\bibinfo{person}{Judith Segal}.}
  \bibinfo{year}{2005}\natexlab{}.
\newblock \showarticletitle{When Software Engineers Met Research Scientists: A
  Case Study}.
\newblock \bibinfo{journal}{\emph{Empirical Software Engineering}}
  (\bibinfo{year}{2005}).
\newblock


\bibitem[\protect\citeauthoryear{Segal}{Segal}{2007}]%
        {segal07_problem}
\bibfield{author}{\bibinfo{person}{Judith Segal}.}
  \bibinfo{year}{2007}\natexlab{}.
\newblock \showarticletitle{Some Problems of Professional End User Developers}.
  In \bibinfo{booktitle}{\emph{Proceedings of the IEEE Symposium on Visual
  Languages and Human-Centric Computing}} \emph{(\bibinfo{series}{VLHCC '07})}.
  \bibinfo{publisher}{IEEE Computer Society}, \bibinfo{address}{Washington, DC,
  USA}, \bibinfo{pages}{111--118}.
\newblock
\showISBNx{0-7695-2987-9}
\urldef\tempurl%
\url{https://doi.org/10.1109/VLHCC.2007.50}
\showDOI{\tempurl}


\bibitem[\protect\citeauthoryear{{Segal}}{{Segal}}{2007}]%
        {segal07_enduser}
\bibfield{author}{\bibinfo{person}{J. {Segal}}.}
  \bibinfo{year}{2007}\natexlab{}.
\newblock \showarticletitle{Some Problems of Professional End User Developers}.
  In \bibinfo{booktitle}{\emph{IEEE Symposium on Visual Languages and
  Human-Centric Computing (VL/HCC 2007)}}. \bibinfo{pages}{111--118}.
\newblock
\showISSN{1943-6092}
\urldef\tempurl%
\url{https://doi.org/10.1109/VLHCC.2007.17}
\showDOI{\tempurl}


\bibitem[\protect\citeauthoryear{Segal and Morris}{Segal and Morris}{2008}]%
        {segal08_ss}
\bibfield{author}{\bibinfo{person}{J. Segal} {and} \bibinfo{person}{C.
  Morris}.} \bibinfo{year}{2008}\natexlab{}.
\newblock \showarticletitle{Developing Scientific Software}.
\newblock \bibinfo{journal}{\emph{IEEE Software}} (\bibinfo{year}{2008}).
\newblock


\bibitem[\protect\citeauthoryear{Torres, Toral, Perales, and Barrero}{Torres
  et~al\mbox{.}}{2011}]%
        {torres2011analysis}
\bibfield{author}{\bibinfo{person}{MR~Martinez Torres}, \bibinfo{person}{SL
  Toral}, \bibinfo{person}{M Perales}, {and} \bibinfo{person}{F Barrero}.}
  \bibinfo{year}{2011}\natexlab{}.
\newblock \showarticletitle{Analysis of the core team role in open source
  communities}. In \bibinfo{booktitle}{\emph{Complex, Intelligent and Software
  Intensive Systems (CISIS), 2011 International Conference on}}. IEEE,
  \bibinfo{pages}{109--114}.
\newblock


\bibitem[\protect\citeauthoryear{Tu, Yu, and Menzies}{Tu et~al\mbox{.}}{2019}]%
        {tu2019better}
\bibfield{author}{\bibinfo{person}{Huy Tu}, \bibinfo{person}{Zhe Yu}, {and}
  \bibinfo{person}{Tim Menzies}.} \bibinfo{year}{2019}\natexlab{}.
\newblock \bibinfo{title}{Better Data Labelling with EMBLEM (and how that
  Impacts Defect Prediction)}.
\newblock
\newblock


\bibitem[\protect\citeauthoryear{van Lamsweerde}{van Lamsweerde}{2009}]%
        {vanLamsweerde2009_requirement}
\bibfield{author}{\bibinfo{person}{Axel van Lamsweerde}.}
  \bibinfo{year}{2009}\natexlab{}.
\newblock \bibinfo{booktitle}{\emph{Reasoning About Alternative Requirements
  Options}}.
\newblock
\showISBNx{978-3-642-02463-4}
\urldef\tempurl%
\url{https://doi.org/10.1007/978-3-642-02463-4_20}
\showDOI{\tempurl}


\bibitem[\protect\citeauthoryear{Vanter, Faulk, Squires, Loh, and Votta}{Vanter
  et~al\mbox{.}}{2009}]%
        {faulk09_secs}
\bibfield{author}{\bibinfo{person}{M.~L. Vanter}, \bibinfo{person}{S. Faulk},
  \bibinfo{person}{S. Squires}, \bibinfo{person}{E. Loh}, {and}
  \bibinfo{person}{L.~G. Votta}.} \bibinfo{year}{2009}\natexlab{}.
\newblock \showarticletitle{Scientific Computing's Productivity Gridlock: How
  Software Engineering Can Help}.
\newblock \bibinfo{journal}{\emph{Computing in Science \& Engineering}}
  (\bibinfo{year}{2009}).
\newblock


\bibitem[\protect\citeauthoryear{{Vasilescu}, {Blincoe}, {Xuan}, {Casalnuovo},
  {Damian}, {Devanbu}, and {Filkov}}{{Vasilescu} et~al\mbox{.}}{2016}]%
        {vasilescu16_limit}
\bibfield{author}{\bibinfo{person}{B. {Vasilescu}}, \bibinfo{person}{K.
  {Blincoe}}, \bibinfo{person}{Q. {Xuan}}, \bibinfo{person}{C. {Casalnuovo}},
  \bibinfo{person}{D. {Damian}}, \bibinfo{person}{P. {Devanbu}}, {and}
  \bibinfo{person}{V. {Filkov}}.} \bibinfo{year}{2016}\natexlab{}.
\newblock \showarticletitle{The Sky Is Not the Limit: Multitasking Across
  GitHub Projects}. In \bibinfo{booktitle}{\emph{2016 IEEE/ACM 38th
  International Conference on Software Engineering (ICSE)}}.
\newblock


\bibitem[\protect\citeauthoryear{Xia, Shu, Shen, and Menzies}{Xia
  et~al\mbox{.}}{2019}]%
        {xia2019sequential}
\bibfield{author}{\bibinfo{person}{Tianpei Xia}, \bibinfo{person}{Rui Shu},
  \bibinfo{person}{Xipeng Shen}, {and} \bibinfo{person}{Tim Menzies}.}
  \bibinfo{year}{2019}\natexlab{}.
\newblock \bibinfo{title}{Sequential Model Optimization for Software Process
  Control}.
\newblock
\newblock


\end{thebibliography}

\end{document}